\documentclass[journal]{vgtc}                %

\ifpdf%
  \pdfoutput=1\relax                   %
  \usepackage{graphicx}                %
  \DeclareGraphicsExtensions{.pdf,.png,.jpg,.jpeg} %
\else%
  \usepackage{graphicx}                %
  \DeclareGraphicsExtensions{.eps}     %
\fi%

\graphicspath{{figures/}{pictures/}{images/}{./}} %

\usepackage{microtype}                 %
\PassOptionsToPackage{warn}{textcomp}  %
\usepackage{textcomp}                  %
\usepackage{mathptmx}                  %
\usepackage{times}                     %
\usepackage{cite}                      %
\usepackage{tabu}                      %
\usepackage{booktabs}                  %

\usepackage{tabularx}

\usepackage{array}
\newcolumntype{L}[1]{>{\raggedright\let\newline\\\arraybackslash\hspace{0pt}}m{#1}}
\newcolumntype{C}[1]{>{\centering\let\newline\\\arraybackslash\hspace{0pt}}m{#1}}
\newcolumntype{R}[1]{>{\raggedleft\let\newline\\\arraybackslash\hspace{0pt}}m{#1}} 

\usepackage{enumitem}
\setlist{nolistsep}

\usepackage{subfigure}

\usepackage{textcomp}
\usepackage{color,soul}
\usepackage[dvipsnames]{xcolor}
\usepackage{todonotes}
\usepackage{soul}

\usepackage[final]{changes}%

\usepackage{wrapfig}

\newcommand{\Room}{\texttt{Rm}}
\newcommand{\RoomOD}{\texttt{RmO}}
\newcommand{\Zoom}{\texttt{Zm}}
\newcommand{\ZoomOD}{\texttt{ZmO}}

\usepackage{setspace}
\onlineid{0}

\vgtccategory{Research}
\vgtcpapertype{evaluation}

\title{Embodied Navigation in Immersive Abstract Data Visualization:\\
Is Overview+Detail or Zooming Better for 3D Scatterplots?}

\author{Yalong Yang, Maxime Cordeil, Johanna Beyer, Tim Dwyer, Kim Marriott and Hanspeter Pfister}
\authorfooter{
\item
 Yalong Yang, Johanna Beyer and Hanspeter Pfister are with the School of Engineering and Applied Sciences, Harvard University, Cambridge, MA, USA. E-mail: \{yalongyang, jbeyer, pfister\}@g.harvard.edu
\item
 Maxime Cordeil, Tim Dwyer and Kim Marriott are with the Department of Human-Centred Computing, Faculty of Information Technology, Monash University, Melbourne, Australia. E-mail: \{max.cordeil, tim.dwyer, kim.marriott\}@monash.edu
}

\shortauthortitle{Biv \MakeLowercase{\textit{et al.}}: Global Illumination for Fun and Profit}

\abstract{
Abstract data has no natural scale and so interactive data visualizations must provide techniques to allow the user to choose their viewpoint and scale.  Such techniques are well established in desktop visualization tools. The two most common techniques are zoom+pan and overview+detail. However, how best to enable the analyst to navigate and view abstract data at different levels of scale in immersive environments has not previously been studied.  We report the findings of the first systematic study of immersive navigation techniques for 3D scatterplots.
We tested four conditions that represent our best attempt to adapt standard 2D navigation techniques to data visualization in an immersive environment while still providing standard immersive navigation techniques through physical movement and teleportation.
We compared room-sized visualization versus a zooming interface, each with and without an overview.  We find significant differences in participants' response times and accuracy for a number of standard visual analysis tasks. Both zoom and overview provide benefits over standard locomotion support alone (i.e., physical movement and pointer teleportation). However, which variation is superior, depends on the task. We obtain a more nuanced understanding of the results by analyzing them in terms of a time-cost model for the different components of navigation: way-finding, travel, number of travel steps, and context switching.

} %


\CCScatlist{ %
 \CCScat{K.6.1}{Management of Computing and Information Systems}%
{Project and People Management}{Life Cycle};
 \CCScat{K.7.m}{The Computing Profession}{Miscellaneous}{Ethics}
}

\teaser{
  \centering
  \includegraphics[width=0.95\linewidth]{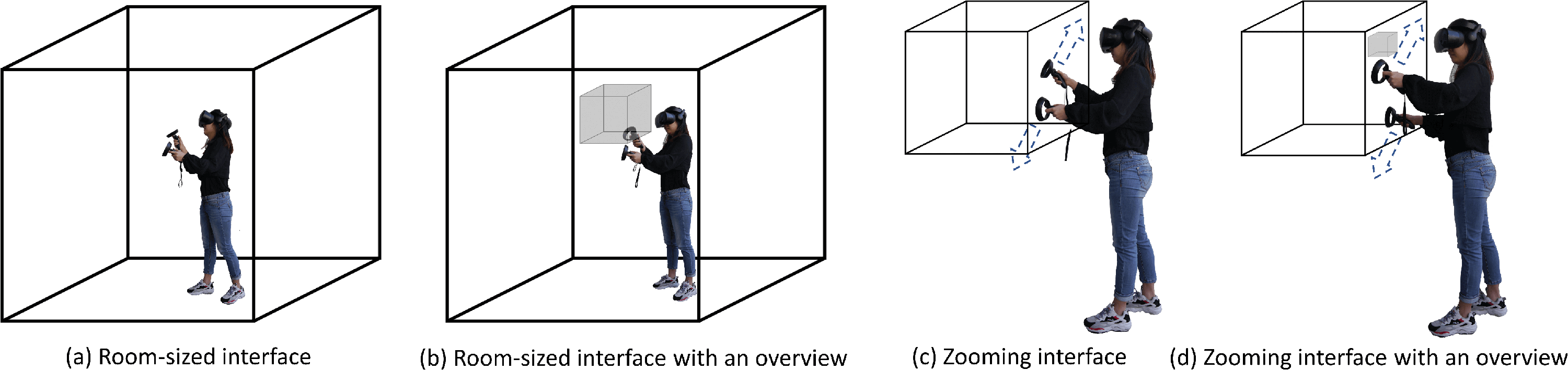}
  \caption{Conditions tested in our user study: a) Room-sized interface (or \Room{}). b) Room-sized interface with an overview (or \RoomOD{}). 
  c) Zooming interface (or \Zoom{}): users zoom in and out with a ``pinch'' gesture. d) Zooming interface with an overview (or \ZoomOD{}).}
  \label{fig:teaser}
  \vspace{1em}
}

\vgtcinsertpkg

\begin{document}

\firstsection{Introduction}
\maketitle
Abstract data has no natural scale.  That is, data that is not based in a physical reference space can be freely re-scaled and viewed from any angle that best supports the analysis task.  However, this freedom also represents a challenge to the design of interactive navigation methods: how do we allow people to move freely through an abstract information space without confusing them?  Designers of desktop visualization tools have grappled with this problem for decades and have evolved sophisticated techniques for navigation.  For example, it is common in data visualizations, such as scatterplots, time-series, and so on, to allow users to zoom in, making a specific region larger to access more detail.  They can then pan around at the new zoom level or zoom out again to reorient themselves in the full dataset.  Another common approach is to provide a minimap window to provide an overview of the whole dataset at all times.  As discussed in Section \ref{sec:related-od-2D}, these methods are well studied and widely accepted in desktop data visualization. 

\added{Recent developments in technology has seen renewed interest in using immersive environments for data visualization, but this requires us to reconsider and possibly adapt navigation methods that have become standard on the desktop.}
As virtual and augmented reality headsets continue to improve~---~e.g.,\ in resolution, field-of-view, tracking stability, and interaction capabilities~---~they present a viable alternative to traditional screens for the purposes of data visualization. A recent study has found a clear benefit to an immersive display of 3D clustered point clouds over more traditional 2D displays~\cite{kraus_impact_2019}.  However, in that study the only way for participants to change their point of view was through physical movement. There are questions about how one might navigate in situations where the participant cannot move around (e.g.,\ they are seated), or where the ideal zoom-level for close inspection of the data is too large to fit in the physically navigable space.  

In Virtual Reality (VR) gaming ``teleporting'' is a standard way to navigate a space that is too large to fully explore through one-to-one scale physical movement.  The equivalent of the desktop minimap overview+detail in VR is World In Miniature (WIM) navigation.
Zooming is also possible in immersive environments through a latched gesture to scale the information space around the user.
However, the effectiveness of such immersive navigation techniques for point cloud data visualization applications in VR has not previously been tested.  The inherent differences between such abstract data visualizations and typical VR worlds, namely, the freedom to move between different scales and viewpoints, makes it unclear whether standard VR navigation methods are sufficient to support analysis tasks.

Our overarching research question, therefore, is: \textbf{how can we best enable the analyst to navigate and view abstract data at different levels of scale in immersive environments?} 
To address this question we first  thoroughly review standard data visualization navigation techniques from desktop environments and the initial work that has begun to provide navigable data spaces in immersive environments (Sec.~\ref{sec:relwork}).  We find that existing literature in immersive data visualization has not fully explored how to adapt navigation techniques for data visualization from the desktop and/or to adapt VR navigation techniques to the application of data visualization.  Therefore, in Section \ref{sec:design}, we look at the space of possible ways in which users can navigate immersive data visualizations and from this, in Sec.~\ref{sec:exp-conditions}, we choose and study four basic navigation possibilities: room-sized data at fixed-scale with physical navigation versus a smaller data display with zooming (dynamic scale), each with and without an overview WIM display. 

Specifically, we compare zooming to overview+detail. We are particularly interested in how to ensure that navigation remains embodied through physical movements and gestures that are as natural as possible \added{\cite{dourish2001action}}.
The contributions of this paper are:
\begin{enumerate}
    \item The first systematic exploration of \emph{overview} and \emph{detail} interaction paradigms in immersive scatterplot visualization, Sec.~\ref{sec:design}.
    
    \item A study comparing four navigation conditions (as above), Sec.~\ref{sec:study} and Sec.~\ref{sec:studyresults}. This finds that: 
    Participants significantly preferred either \emph{zoom} or \emph{overview}  over standard locomotion support alone. Adding an \emph{overview} also improved accuracy in some tasks. Room-sized egocentric views were generally faster in our study.
    
    \item A more nuanced interpretation of our study results using a \emph{navigation time-cost model},  Sec.~\ref{sec:timecostmodel}.
    
    \item Recommendations for use of \emph{zoom} and \emph{overview} in combination with physical navigation for immersive data visualization, Sec.~\ref{sec:conclusion}.
\end{enumerate}

\section{Related Work}
\label{sec:relwork}

\subsection{Navigation in 2D Visualizations}
\label{sec:related-od-2D}
A foundation of visual data exploration is the Shneiderman \emph{Information Seeking Mantra} of ``Overview first, filtering and details on demand'' \cite{shneiderman1996eyes}.
Several paradigms have emerged to support navigation 
following this mantra, such as overview+detail, focus+context \cite{mackinlay1991perspective} and zooming \cite{card1999readings}. Many 2D visualization techniques provide such interactions, such as Google Map (zooming), PowerPoint (overview+detail), and fish-eye views (focus+context).
In an extensive literature review, Cockburn \textit{et al.}~\cite{cockburn2009review} conclude that each technique has issues~---~zooming impacts memory~\cite{cockburn2004comparing};
focus+context distorts the information space; and overview + detail views also make it difficult for users to relate information between the overview and the detailed view \cite{hornbaek2002navigation, baudisch2002keeping}.

Navigation has also been studied on mobile displays.
Burigat \textit{et al.}~\cite{burigat2013effectiveness} found that interactive overviews were 
particularly advantageous on small-screen mobile devices for map navigation tasks~\cite{burigat2008map,burigat2013effectiveness}. However, in the context of information visualization, adding an overview to a zoomable scatterplot on a PDA was found to be ineffective and detrimental compared to a zoomable user interface only~\cite{buring2006usability}.

With screens and projectors becoming cheaper, larger and easier to tile, researchers have explored the benefits of large display sizes for navigating visualizations. Ball and North \cite{ball2005effects} studied the effect of display size on 2D navigation of abstract data spaces and found that larger displays and physical navigation
outperformed smaller displays.
They also found that virtual navigation (i.e., pan and zoom) was more frustrating than physically navigating a large tiled-display. 
\deleted{Later,} Ball and North \cite{ball2008effects} \added{also} investigated the reason for improved performance of information space navigation on large displays and found that physical movements were the main factor, compared to field of view. 
\deleted{Later,} Another study by Ronne and Hornbaek \cite{ronne2011sizing} investigated focus+context, overview+detail and zooming on varying display sizes. They found that small size display was least efficient, but found no differences between medium and large screens and that overview+detail performed the best.

In our study, we investigate the use of overview+detail and \emph{continuous} zoom, as they are most suited to our tasks. We do not consider focus+context as it is not well suited to our tasks (i.e., counting points and evaluating distances between pairs of points), and spatial distortion would introduce misinterpretation of distances.

\subsection{Navigation in VR Environments}
\label{sec:related-iv-nav}

Research in 3D User Interfaces (3DUI) and Virtual Reality has explored different ways for users to navigate 3D immersive scenes. Laviola et al.~\cite[Ch.~8]{laviola20173d} provide a taxonomy of these techniques.
They describe four navigation metaphors~---~walking, steering, selection-based travel and manipulation based travel. 

Walking in VR can be either \emph{real walking} (i.e., the user walks with a VR headset in the real world) or an assisted form of walking (e.g., using a treadmill~\cite{fung2006treadmill} or more dedicated devices to emulate walking in place~\cite{iwata2001gait}). While those walking devices provide a realistic sensation of walking in the virtual world, they are bulky and not affordable for a wide audience. 
\added{Abtahi et al~\cite{abtahi_2019_Giant} recently investigated mechanisms to increase users’ perceived walking speed for moving in a large virtual environment. However, it is not trivial on how to apply their techniques for visualization navigation.}
In our research, we focus on physical walking available through HMD VR devices, however restricted to a common office size area. Walking has been found to be more efficient than using a joystick for orientation in VR~\cite{chance1998locomotion}, and has generally been found to be beneficial for perception and navigation tasks compared to controller-based interactions~\cite{ruddle2009benefits, ruddle2011walking, lages_move_2018}. Walking has also been proven to be more efficient than standing at a desk for tasks involving visual search and recalling item locations in space~\cite{radle2013effect}. 

In contrast to walking, steering metaphors allow the user to navigate the space without physically movement. 
They are usually gaze- and/or controller-based.
Teleportation is a very common locomotion technique which consists of casting a ray from a tracked controller to the desired location~\cite{mine1995virtual} and a button push to travel there.
Bowman et al.~\cite{bowman1997travel} found that hand-based steering was more efficient than gaze-based travel, but that instant teleportation (with no transition between initial and final location) was detrimental for orientation compared to  continuous viewpoint change. Laviola et al.~\cite{laviola20173d} argue that continuous uncontrolled movements of the viewpoint induce cybersickness, but that it can be mitigated if the transition is very fast.
Modern teleportation techniques use fade or blinking metaphors where the screen transitions to black before the move and is restored at the destination position. %
In all the conditions of our experiment we allow this type of teleportation to allow the user to fast track travel in the virtual environment if needed.

An alternative VR navigation technique is WIM~\cite{stoakley1995virtual,mine1995isaac}, which introduces a ``minimap'' to allow the user to teleport by selecting a target position in the miniature. 
Recent refinements offer scalable, scrollable WIMs (SSWIMs)~\cite{wingrave2006overcoming}, or support for selecting the optimal viewpoint in dense occluded scenes~\cite{trueba2009complexity}. We designed an embodied placement for WIM, and two different teleportation methods with WIM.

In summary the effects of spatial memory, cybersickness and overall usability of such techniques have been studied in general purpose VR applications. In the context of abstract visualization their relative drawbacks and benefits are unclear. %

\subsection{Immersive Analytics and Space Navigation}
\label{sec:related-immersive-data-vis-nav}

3D scatterplots have received significant attention in Immersive Analytics research ~\cite{bach_hologram_2018,cordeil_imaxes:_2017,fonnet2018axes,kraus_impact_2019,prouzeau_scaptics_2019}. Raja et al.~\cite{raja2004exploring} found that using body motion and head-tracking in a CAVE-like environment was beneficial compared to non-immersive environments for low-level scatterplot tasks (i.e., cluster detection, distance estimation, point value estimation and outlier detection). Kraus et al.~\cite{kraus_impact_2019} explored the effect of immersion for identifying clusters with scatterplots. They explored a 2D scatterplot matrix and 3D on-screen, and immersive table-size and room-size spaces. Overall their found that the VR conditions were well suited to the task. However they also pointed out that their room-size condition may not have been optimal due to the lack of an overview.

Body movements have been observed in Immersive Analytics studies involving 3D scatterplots, which may indicate potential benefits for data exploration and presentation. For example, Prouzeau et al.~\cite{prouzeau_scaptics_2019} observed that some participants tended to put their heads inside 3D scatterplots to explore hidden features; Batch et al.~\cite{batch2019there} observed participants organize their visualizations in a gallery-style setup and walk through them to report their findings. Simpson et al.~\cite{simpson2017take} informally studied walking versus rotating in place for navigating a 3D scatterplot, and their preliminary result indicated that participants with low spatial memory were more efficient in the walking condition.

Other alternatives to walking in a room-size environment have also been explored. Filhio et al.~\cite{filho_virtualdesk:_2018} found that a seated VR setup for 3D scatterplots performed better than on a desktop screen. Satriadi et al.~\cite{satriadi2019augmented} tested different hand gesture interactions (user standing still) to pan and zoom 2.5D maps in Augmented Reality, but the visualizations were  within a fixed 2D viewport in the 3D space. 

Flying is a steering metaphor that has been used as an alternative to walking to obtain a detailed view of an immersive visualization. With flying, benefits of immersion were found compared to a 2D view for distance evaluation tasks in scatterplots~\cite{wagner_filho_immersive_2018}. Sorger et al.~\cite{sorger2019immersive} designed an overview+detail immersive 3D graph navigation metaphor that uses flying to gain an overview, and a tracked controller-based teleport to a selected node position to obtain details from a node-centric perspective. The main takeaway of their informal expert-study is that overviews were perceived as important to keep the user oriented in the graph.

To our knowledge, WIMs have been underexplored in immersive data visualization. Nam et al.~\cite{nam_worlds--wedges_2019} introduced the Worlds-in-Wedges technique that combines multiple virtual environments simultaneously to support context switching for forest visualization. The study of Drogemuller et al.~\cite{drogemuller_examining_2020} was the first to formally evaluate one and two-handed flying vs.\ teleportation and WIM in the context of large immersive 3D node-link  diagrams\cite{drogemuller_examining_2020}. They found that the flying methods were faster and preferred compared to teleportation and WIMs, for node finding and navigation tasks in the 3D graph. However, they did not test the condition where the user only walks around the room without instrumented metaphors for overview+detail.

Finally, scaling the visualization as a 3D object in VR via bimanual interaction has been used in Immersive Analytics systems \cite{hurter2018fiberclay, wagner_filho_evaluating_2019} but not formally studied against other VR navigation techniques.
In summary, there is evidence that overview+detail and zooming are beneficial for navigating 2D visualizations. 
However, their relative benefit in the context of immersive 3D scatterplots is still underexplored.

\section{Study Rationale and Designs}
\label{sec:design}
\begin{figure}
	\centering
	\includegraphics[width=.8\columnwidth]{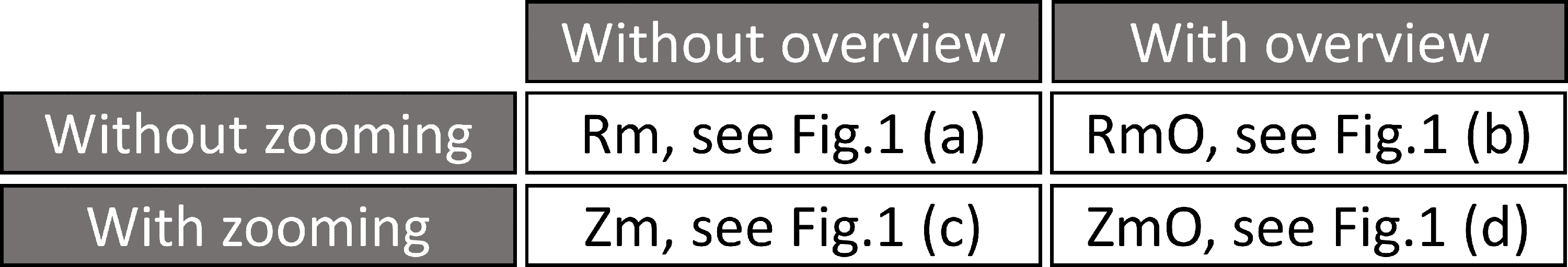}
	\caption{
	Our study compares two main effects, leading to four conditions.
	}
	\label{fig:study-rationale}
\end{figure}

Our study is intended to address the gaps in the literature described above in terms of how to carry over well understood 2D navigation techniques to immersive visualizations (see Fig.~\ref{fig:study-rationale}). 
\added{We decided to not include 2D interfaces for 3D scatterplots as testing conditions, as the 2D alternatives (i.e., scatterplot matrix and 3D scatterplot on a 2D screen) have been tested to be less effective than representing them as 3D scatterplots in an immersive environment~\cite{kraus_impact_2019}.}

We focus on adapting zoom and overview+detail techniques.
We did not include focus+context in our study, as it has been shown that such spatial distortion techniques are not effective for large 2D display spaces~\cite{ronne2011sizing}, and we expect similar results in the immersive environments.
However, formal confirmation of this is worthwhile future work.
Adapting overview+detail and zooming techniques to immersive environments from 2D interfaces requires many design decisions. In the following, we describe our design considerations and choices. 

\subsection{Overview+Detail}
\label{sec:overview_plus_detail}
The idea of the overview+detail is to provide two separate display spaces representing the same information but at different scales. The basic design of our overview is straightforward: the whole information space in which the user is standing is represented in a cube -- with all data glyphs scaled down appropriately.  
However, three essential design choices remain unanswered: \emph{placement of the overview; Point of View (PoV) indicator in the overview; and overview teleportation.}

\vspace{0.3em}
\noindent\textbf{Placement of the overview:} 
In 2D display environments, the overview is commonly placed at a global fixed position relative to the display: either at a corner of the detail view (e.g., some on-line maps) or outside the detail view (e.g., some text editors, PowerPoint).
When using the overview on a 2D display, the user's Field of View (FoV) is consistent and can cover the full display space at all times. Thus, the user can easily access the overview at any time in this placement.
Unlike 2D displays, in an immersive environment, the user is expected to perform more body movement. As a result, the user's FoV is constantly changing. If the overview is placed at a fixed global position in immersive environments, the user may forget its location or it may not be reachable when required.  In general, the overview needs to be easily accessible so it can be brought into focus for close inspection, but by default it should not occlude the users' view of the main scene.  A logical design choice is to place the overview somewhere at the periphery of the users' view by default, but allow the user to grab it with the controller to bring it up for close inspection.  But there remains the question of what is the best default location relative to the user.

Our first idea was to place the overview at a fixed position relative to the FoV, i.e., the overview follows the user's movement, e.g., to always appear at the bottom right corner of the user's FoV. 
We tested it with two participants; both of them reported the overview to be distracting and cluttered. They also found it difficult to access (grab) the overview.

\begin{figure}
    \vspace{0.5em}
	\centering
	\includegraphics[width=\columnwidth]{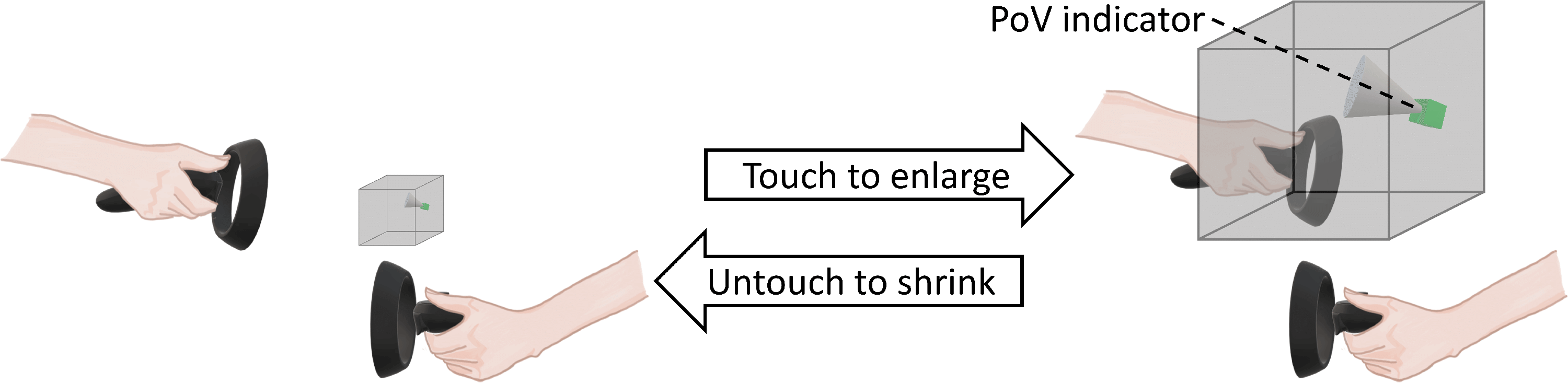}
	\caption{Enlarging and shrinking the overview with arm movement. }
	\label{fig:overview-enlarge}
\end{figure}

We then attached the overview to the user's off-hand controller.
This was inspired by WIM where the miniature is associated with a tracked physical board~\cite{stoakley1995virtual,mine1995isaac}, and Mine et al.~\cite{mine_moving_1997} who attached widgets to virtual hands in VR.
To further reduce visual clutter, we shrink the overview by default and allow the user to enlarge the overview by touching it with the other controller. 
This embodied design allows users to enlarge or shrink the overview by simple hand movements (see Fig.~\ref{fig:overview-enlarge}).
We tested it with another participant, and explicitly asked if she felt the overview caused visual clutter, was distracting, or difficult to access. The participant was highly positive about the new design.

\begin{figure}
    \vspace{0.5em}
	\centering
	\includegraphics[width=0.8\columnwidth]{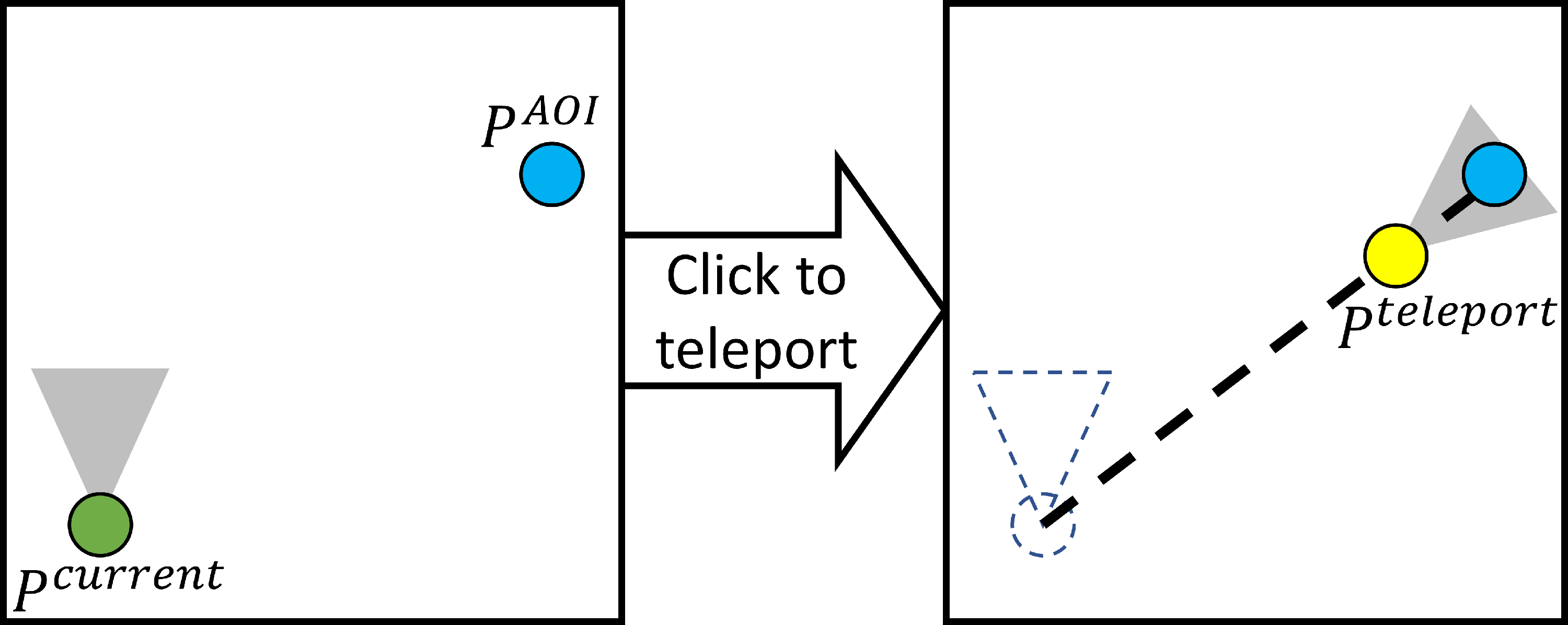}
	\caption{Demonstration of point-and-click teleportation. $p^{current}$ (the green dot) is the current position of the user; $p^{AoI}$ (the blue dot) is the position of Area of Interest (AoI) \added{and the clicked position}; and $p^{teleport}$ (the yellow dot) is the position the user will teleport to.}
	\vspace{-0.5em}
	\label{fig:click_to_teleport}
\end{figure}

\vspace{0.3em}
\noindent\textbf{PoV indicator in the overview:}
In 2D overview+detail interfaces, a FoV box is used to indicate which portion of the overview is presented in the detail view.
Similarly, in WIM, it is also important to indicate the view position, but in addition the direction the user is facing~\cite{stoakley1995virtual}. 
Following their designs, we use a cube to represent the tracked headset. The position and rotation of the cube are synchronized with the headset in real-time. We also use a semi-transparent cone attached to the cube to represent the user's view direction (see Fig.~\ref{fig:overview-enlarge}). 

\vspace{0.3em}
\noindent\textbf{Overview teleportation:}
Tight coupling between the overview and detail views is  standard in 2D interfaces (e.g.,~\cite{hornbaek2002navigation,woodburn_interactive_2019}). 
The aim is to allow the user to change the portion of the scene presented in the detail view by interacting with the overview.
There are two widely used implementations: \emph{drag-and-drop} and \emph{point-and-click} interactions. 
Changing the detail view in 2D interfaces is equivalent to changing the viewing point and direction in immersive environments. 
We adapted the 2D interactions for immersive environments: 

\emph{Drag-and-drop teleportation:} 
On 2D interfaces, the user can select the FoV box in the overview then drag-and-drop it at a new position in the overview. The detail view will then switch to the new position.
In immersive environments, WIM allows participants \emph{``pick themselves up''}, i.e., the user can pick up the PoV indicator and drag it to a new position in the overview to teleport to the dropped location in the detail view. We implemented the same mechanism in our visualizations.

\emph{Point-and-click teleportation:}
The user can also directly choose the destination and explicitly trigger a command to change the detail view.
On 2D interfaces, the operation involves pointing a cursor at the destination position in the overview and then clicking to translate the detail view.
In immersive environments, in addition to changing the position of the user, we also need to adjust the user's orientation. We implemented a mechanism that will determine the user's position and orientation (see Fig.~\ref{fig:click_to_teleport}). Basically, the teleport target position is on the straight line connecting the current position and the Area of Interest (AoI), but slightly away from the AoI to ensure it is within the user's FoV. The orientation is the direction of this straight line.

Drag-and-drop teleportation is expected to give users more control of their position and orientation and possibly give them a better estimation of their position and orientation after the teleportation.
However, this multi-step operation is not welcomed by all users~\cite{kumar_browsing_1997}. 
Point-and-click teleportation requires less steps, but may increase the gap between the expected and actual teleported position and rotation.
Like many other 2D interfaces, we include both mechanisms in our visualizations, to let users choose their preferred option.

\begin{figure}
	\centering
	\includegraphics[width=0.75\columnwidth]{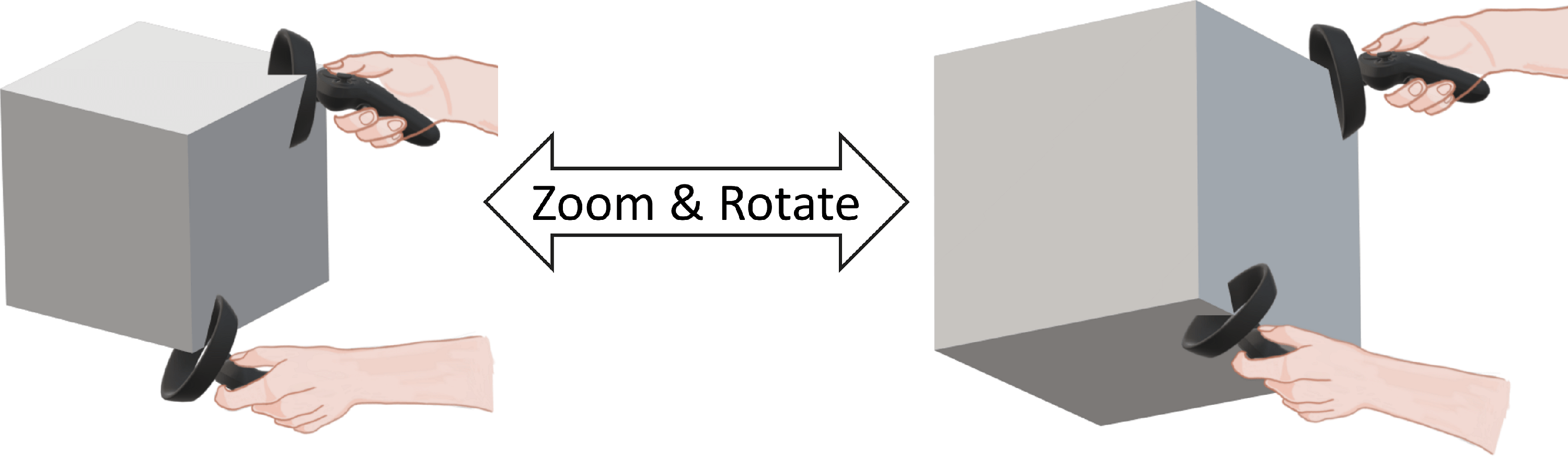}
	\caption{The gesture to zoom and rotate simultaneously. From left to right: scale up the object and rotate clockwise; from right to left: scale down the object and rotate anti-clockwise.}
	\label{fig:zoom_and_rotate}
\end{figure} 

\subsection{Zooming}
\label{sec:zooming}
Although 2D desktop interfaces usually use the scroll-to-zoom metaphor, pinch-to-zoom is the standard zooming gesture on most multi-touch devices.
The idea of pinch-to-zoom is to re-scale
according to the distance between two touch points~\cite{hinckley_interaction_1998}.
Immersive systems can naturally be considered as multi-touch systems as they are capable of tracking at least two hand-held controllers.
Therefore, we use the pinch-to-zoom gesture in our visualizations.

The naive implementation of pinch-to-zoom in immersive environments (i.e., only scaling the size of the object) shifts the positions of the controllers relative to the object. 
As a result, inconsistency is introduced between the interaction and the displayed information, which confuses the user. 
To address this issue, we integrate rotation into the same gesture, i.e., the rotation of the manipulated object is based on the direction between the two touch points while zooming. 
Additionally, we apply adjusted 3D translation to the object to ensure the positions of the controllers relative to the zoomable object are preserved in the interaction (i.e., \emph{latching}).
The simultaneous zoom and rotate gesture in the 3D immersive environment is demonstrated in Fig.~\ref{fig:zoom_and_rotate}.
A similar concept is also widely used on 2D multi-touch zooming interfaces, e.g., photo editing interfaces on mobile phones.

\begin{figure*}
	\centering
	\includegraphics[width=\textwidth]{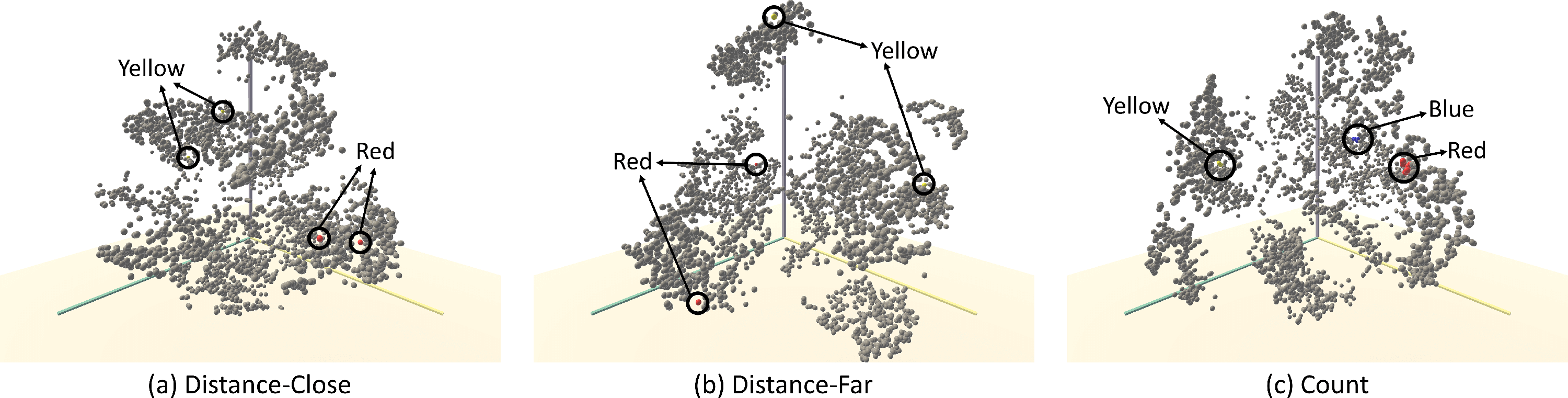}
	\caption{Example stimuli for three task conditions. The labels are only for demonstration purposes. %
	(a,b) Distance: participants had to estimate which pair of colored points (yellow or red) has the larger spatial distance. The distance within each pair varies in (a) and (b) to be relatively close and far respectively.
	(c) Count: participants had to find which group of colored points (yellow, red or blue) has the largest number of points.}
	\label{fig:study-tasks}
\end{figure*}

\section{User Study}
\label{sec:study}
We pre-registered our user study at \url{https://osf.io/ycz5x}. We also include results of statistical tests as supplementary materials. Test conditions are also demonstrated in the supplemental video.

\subsection{Experimental Conditions}
\label{sec:exp-conditions}
In addition to the two main effects discussed in Sec.~\ref{sec:design}, we summarize the characteristics and parameters of test conditions in Tab.~\ref{tab:design-space}.

\vspace{0.5em}
\noindent\textbf{\Room{}:}
To allow the user to explore finely-detailed data, we take advantage of the large display space in immersive environments and create a room-sized visualization, see Fig.~\ref{fig:teaser}(a).
Room-sized design in immersive environments is considered to be more immersive than table-sized visualizations~\cite{kraus_impact_2019,yalong_yang_maps_2018}.
A few types of room-sized visualizations have been explored, e.g., node-link diagrams~\cite{kwon_study_2016}, egocentric globes~\cite{yalong_yang_maps_2018}, and scatterplots~\cite{kraus_impact_2019,prouzeau_scaptics_2019}.
Room-sized visualizations are expected to scale better for representing finely detailed data. However, only sub-parts of the visualization can be seen by the viewer at one time, so that the user may lose context during exploration. 

We use a 3D cube of 2$\times$2$\times$2 meters as the display space of the room-sized visualization. 
Manipulating a large display space in an immersive environment is likely to introduce strong motion sickness~\cite{yalong_yang_maps_2018}. We therefore keep the position and rotation of the display space fixed. 
Participants can freely move by walking (a few steps) within the space as well as use pointer teleportation, i.e., the user can use a pointer to choose the destination on the floor and then click the controller button to teleport to the destination in the immersive environment. Such pointer teleportation is now a standard locomotion mechanism in major VR platforms (like Oculus, SteamVR and Windows Mixed Reality). We enable this standard functionality in all our tested conditions. This is the only support for navigation in this condition. 

\begin{table}
	\centering
    \includegraphics[width=\columnwidth]{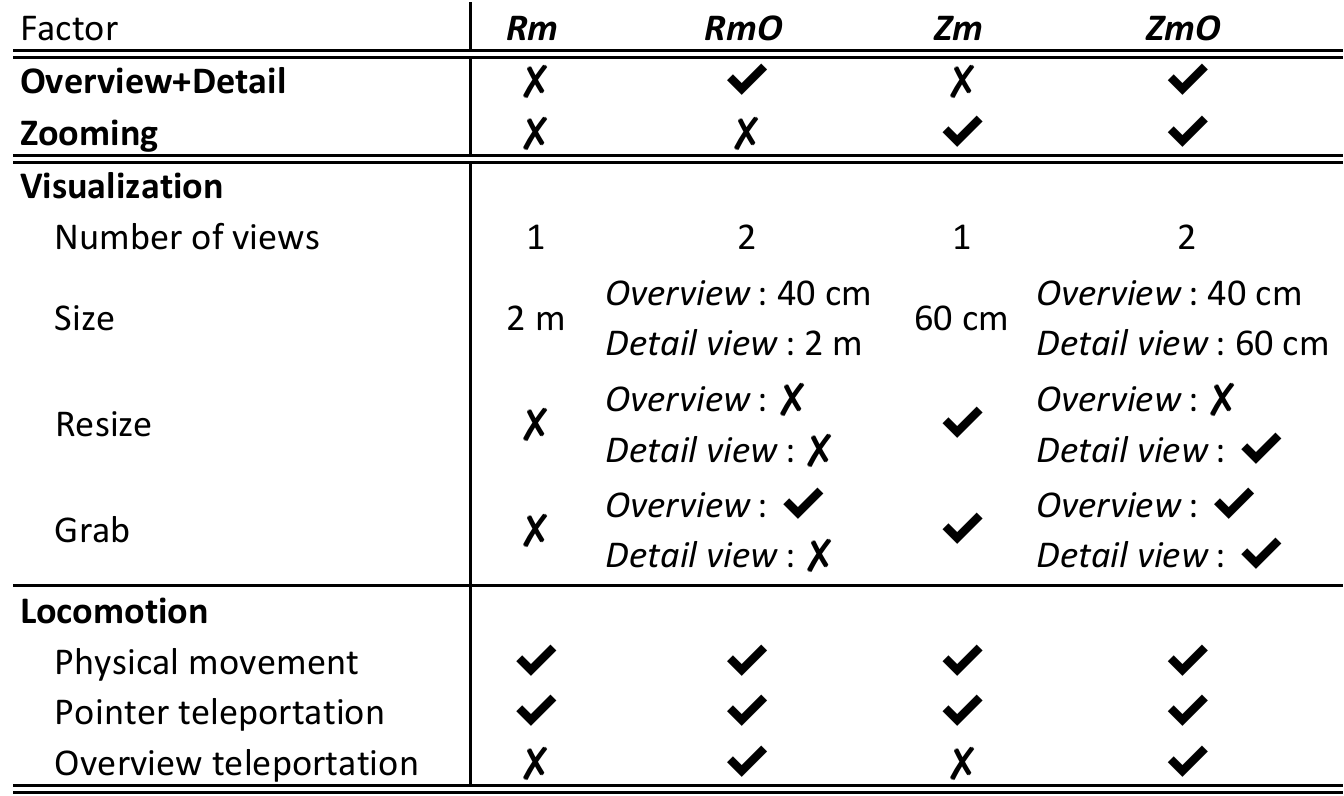}
    \caption{Characteristics and parameters of test visualizations.}
    \label{tab:design-space}
    \vspace{-2em}
\end{table}

\vspace{0.5em}
\noindent\textbf{\RoomOD{}:}
In this condition, in addition to \Room{}, we provide an overview of the display space, see Fig.~\ref{fig:teaser}(b).
The detail view of \RoomOD{} is the same as \Room{}. The overview is a 3D cube of 40$\times$40$\times$40 centimeters when it is enlarged (see Fig.~\ref{fig:overview-enlarge}). 
The default size of the overview is 10$\times$10$\times$10 centimeters. 
Participants can grab the overview by moving a hand-held controller inside and hold the trigger button. The overview will then attach to this controller. Moving and rotating the controller will also move and rotate the overview.
We did not allow participants to resize the overview, as it would increase the complexity of the interactions. 
Participants can teleport using the overview as discussed in Sec.~\ref{sec:overview_plus_detail}.

\vspace{0.5em}
\noindent\textbf{\Zoom{}:}
The visualization is initially table-sized (60$\times$60$\times$60 centimeters), and participants can use the pinch-to-zoom gesture (described in Sec.\ \ref{sec:zooming}) for resizing, see Fig.~\ref{fig:teaser}(c). 
The small initial size allows participants to have an overview of the information first, which is recognized as the standard analytic workflow.
We also allow participants to grab the view and manipulate it with a hand-held controller.

\vspace{0.5em}
\noindent\textbf{\ZoomOD{}:}
In this condition, in addition to \Zoom{}, we provide an overview of the zooming view, see Fig.~\ref{fig:teaser}(d). 
The detail view of \ZoomOD{} is the same as \Zoom{} (with an initial size of 60$\times$60$\times$60 centimeters), and the overview is the same as it is in \RoomOD{} (with an enlarged size of 40$\times$40$\times$40 centimeters and default size of 10$\times$10$\times$10 centimeters).

\subsection{Experimental Setup}
We used a Samsung Odyssey virtual reality headset with a 110\textdegree~field of view, 2160$\times$1200 pixels resolution and 90~Hz refresh rate. The PC was equipped with an Intel i7-9750H 2.6~GHz processor and NVIDIA GeForce RTX 2060 graphics card. The study took place in a space with approximate 2.5$\times$2.5 meters (6.25~m$^2$). 

Following feedback from our pilot user study and as a common practice for reducing motion sickness in VR~\cite{mcgill_i_2017,yalong_yang_maps_2018}, we created an external reference frame, which is a 4$\times$4 meters virtual floor.

\Room{} and the detail view of \RoomOD{} shared the same center of the physical room. Their orientations were identical in the whole study. 
\Zoom{} and the detail view of \ZoomOD{} were placed 50 centimeters in front of the user's head position and 50 centimeters below it. 
This setup allows users to easily reach the visualization and keep the full visualization within their FoV at the beginning of each trial. 
We repositioned and resized the visualization at the beginning of every trial. We also asked participants to move back to the center before the start of every trial.

\subsection{Data}
We used MNIST~\cite{lecun_gradient-based_1998}, a \emph{real-world} handwritten digits database to generate point cloud data. 
For each data set, we first randomly sampled 5,000 images as data points 
, and then used t-SNE~\cite{maaten_visualizing_2008} to calculate their projected 3D positions (i.e., 5,000 points per scatterplot). 
The t-SNE technique projects high-dimensional data into two or three dimensions. 
In our case, we used TensorFlow's projection tool~\cite{embedded-projector} as the implementation of t-SNE to project 784 dimensions (28 $\times$~28 pixels) per image to three dimensions. We kept the default parameters and executed 800 iterations per data set, which has been tested to be sufficient for getting a stable layout. 
We used different data sets for all trials.
In total, we generated 60 data sets (4 for visualization training trials, 16 for task training trials, and 40 for study trials). 
All points are colored gray, except the red-, yellow-, and blue-colored targets in tasks.
Points in the scatterplot were rendered as spheres with 3 cm diameter in \Room{} and the detail view of \RoomOD{}. The size of the points was proportional in other representations. Sample data is shown in Fig.~\ref{fig:study-tasks}.

\subsection{Tasks}
Sarikaya and Gleicher proposed a task taxonomy for scatterplots. This identified three types of high-level tasks: 
object-centric, browsing, and aggregate-level~\cite{sarikaya_scatterplots_2018}.
Instead of investigating fine detailed information, browsing and aggregate-level tasks are looking at general patterns, trends or correlations. 
These types of tasks require relatively less navigation effort, and participants preferred small-sized display spaces for some of these tasks~\cite{kraus_impact_2019}. 
To better understand the navigation performance,
we intended to select tasks that require relatively significant navigation efforts. 
For our user study, we chose two object-centric tasks that require the participant to navigate to the fine detailed data.

\vspace{0.5em}
\noindent\textbf{Distance:} \emph{Which of the point pairs are further away from each other: the red pair or the yellow pair?} 
Comparing distance is representative of a variety of low-level visualization tasks, e.g., identifying outliers and clusters. These tasks are also essential parts of high-level analysis processes, 
e.g., identifying misclassified cases and understanding their spatial correlation with different nearby classes in the embedding representation of a machine-learning system.
Variations of this task have been studied in most of the studies we reviewed~\cite{kraus_impact_2019,bach_hologram_2018,prouzeau_scaptics_2019,wagner_filho_immersive_2018,wei_evaluating_2020}. 
Among them, we directly adopted this task from Bach et al~\cite{bach_hologram_2018}. 
Similar to their study, we had two target pairs: one pair was colored red, and the other pair was colored yellow (see Fig.~\ref{fig:study-tasks} (a,b)). The point cloud was dense yet sparse enough that the target points could be identified without the need for interactions other than changing the viewing direction. 
In all conditions, the participants had to first search for the targets and then compare their 3D distance by moving/teleporting around the space and/or rotating the visualization when available (see Tab.~\ref{tab:design-space}). Participants needed to choose from two choices: red or yellow.

Whether the two points in a pair can be both presented in the FoV can be a key factor affecting the performance in immersive visualizations~\cite{yalong_yang_maps_2018}. 
We investigated this factor by creating two categories of distance: \textbf{\textit{Close}} and \textbf{\textit{Far}}. 
In \textbf{\textit{Close}}, the larger distance of the pair was controlled to be 25\% of side length of the view (e.g., 0.5 meters in \Room).
In \textbf{\textit{Far}}, we controlled this parameter to be 75\% (e.g., 1.5 meters in \Room).
We also controlled the difference between the distances of the two pairs to be 10\%. 
We expect a small difference can potentially encourage participants to verify their answers from different viewing positions and directions and thus increase the number of navigations.
For the same reason, we placed each pair far away from the other pair (i.e., they had a distance of 75\% of the side length of the view, for example, 1.5 meters in \Room{}).
We developed an automatic strategy to select the points to meet all controlling requirements. We first repeatedly randomly select a pair of points from the 5000 points in a data set until the pair meets the distance requirement for \textbf{\textit{Close}} or \textbf{\textit{Far}} and then keep randomly selecting the other pair until all other requirements are met.

\vspace{0.5em}
\noindent\textbf{Count:} \emph{Which group has the largest number of points: the red group, the yellow group, or the blue group?} This task is essential for processes that require the understanding of numerosity, e.g., counting the number of misclassified cases or cases with specific properties for a classification system. 
Again, variations of this task have been studied in some of the studies we reviewed~\cite{bach_hologram_2018,wagner_filho_immersive_2018,wei_evaluating_2020}. 
We created three groups of points, which were colored in red, yellow, and blue (see Fig.~\ref{fig:study-tasks} (c)). The points in a group were close to each other to form a small cluster. Points could partially overlap with each other, but we made sure that the number of points is unambiguous. 
The participant has to first search for the groups and then sequentially get close to each group to count the number of points within that group. 
In all conditions, participants can move or teleport around the space. In \Zoom{} and \ZoomOD{}, participants also need to enlarge the view to count the points. Participants are not able to count the points in the overview in \RoomOD{} and \ZoomOD{} due to its small size. Participants need to choose from three choices: red, yellow, or blue.

The number of points in a group varied from 5 to 10. Again, to increase the potential number of navigation steps needed to complete this task, we placed groups far apart from each other (approximate 50\% of the side length of the view, that is, e.g., one meter in \Room{}). 
Unlike in the Distance tasks where we used an automatic process to select targets, in the Count task, such method may produce ambiguous overlapping cases. Instead, we manually selected groups of points for each trial.

\subsection{Participants}
We recruited 20 participants (14 females and six males) from Harvard University. 
All had a normal or corrected-to-normal vision, were right-handed, and all were college students.
All participants were within the age range of 20$-$30.
VR experience varied: three participants had no experience before this user study; ten participants had 0$-$5 hours experience; four participants had 5$-$20 hours experience, and three participants had more than 20 hours of experience. Most of our participants do not play computer/video/mobile games frequently: 17 participants reported they played less than 2 hours of games per week, and the other three participants played 2$-$5 hours per week.
We provided a \$20 gift card as compensation for each participant.

\subsection{Design and Procedures}
The experiment followed a full-factorial within-subject design. 
We used a Latin square (4 groups) to balance the visualizations, but kept the ordering of tasks consistent: \emph{Distance} then \emph{Count}. The experiment lasted 1.5 hours on average. Each participant completed 40 study trials:
\noindent 4 VR conditions $\times$ (3 \emph{Distance-Close} $+$ 3 \emph{Distance-Far} $+$ 4 \emph{Count})

Participants were first given a brief introduction to the experiment and VR headset. After putting on the VR headset, we asked them to adjust it to see the sample text in front of them clearly. We then conducted a general VR training session to teach participants how to move in VR space and how to manipulate a virtual object. 
First, we asked participants to move for a certain distance physically. Then we told them to touch the touchpad to enable the pointer and click the touchpad to teleport. 
We asked participants to get familiar with the pointer teleportation with a few more teleportations. At the final stage of the pointer teleportation training, we asked them to teleport to a place marked with a green circle on the floor. 
We then asked the participants to grab a green cube by putting the controller inside the cube and holding the trigger button. The participants finished the training session by placing the green cube at a new indicated position with a specific rotation. All participants completed the training and reported that they were familiar with the instructed interactions. The training session took around 5 minutes.

We conducted a visualization training session every time a participant encountered a visualization condition for the first time.
In the training session, we introduced the available interactions and asked participants to get familiar with them with no time limit. Each condition (visualization$\times$task) started with 2 training trials followed by timed study trials. Before each trial, we re-positioned participants to the room's center and faced them in a consistent direction. 
In the training trials, participants were not informed about specific strategies for completing the task but were encouraged to explore their own strategies. The correct answer was presented to them after they had selected an answer in the training trials. If a participant answered incorrectly, we asked the them to review the training trial and verify their strategies.

After each task, participants were asked to fill in a questionnaire regarding their strategies in each visualization, their subjective ratings of confidence, mental and physical demands of each visualization, and  to rank the visualizations based on their preference. We had a 5-minute break between two tasks. After completing two tasks, participants were asked to fill another survey rating the overall usability and discussing the pros and cons of each visualization. The demographic information was collected at the end of the user study. The questionnaire listed visualizations in the same order as presented in the experiment.

\subsection{Measures}
We measured \emph{time} from the first rendering of the visualization to a double-click of the controller trigger. After the double-click, the visualization was replaced by a multiple-choice panel with task description and options. Participants' choice was compared to the correct answer for their \emph{accuracy}.
We recorded the \emph{position} and \emph{rotation} of the headset, controllers, and visualizations every 0.2 seconds. 
We also recorded the number of different interactions participants conducted in each study trial, including \emph{teleportation} and \emph{zooming}. We also collected the \emph{overview usage percentage}, which is the percentage of time the participant was looking at the overview of each study trial. The size of both \Zoom{} and \ZoomOD{} were also collected every 0.2 seconds.
\added{In the pilot study, we also asked participants to report the level of motion sickness they experienced in each condition. All participants reported the minimal level of motion sickness for all conditions. This could be because that the participant's FoV was not fully occupied at any time, and the participant could easily access the visual reference (the floor). Thus, we decided to not record the motion sickness level in the formal study.}

\subsection{Statistical Analysis}
For dependent variables or their transformed values that can meet the normality assumption, we used \emph{linear mixed modeling} to evaluate the effect of independent variables on the dependent variables~\cite{Bates2015}. 
Compared to repeated measure ANOVA, linear mixed modeling is capable of modeling more than two levels of independent variables and does not have the constraint of sphericity~\cite[Ch.\ 13]{field2012discovering}.
We modeled all independent variables and their interactions as fixed effects. A within-subject design with random intercepts was used for all models. 
We evaluated the significance of the inclusion of an independent variable or interaction terms using log-likelihood ratio. 
We then performed Tukey's HSD post-hoc tests for pair-wise comparisons using the least square means~\cite{Lenth2016}. 
We used predicted vs. residual and Q~---~Q plots to graphically evaluate the homoscedasticity and normality of the Pearson residuals respectively. 
For other dependent variables that cannot meet the normality assumption, we used a \emph{Friedman} test to evaluate the effect of the independent variable, as well as a Wilcoxon-Nemenyi-McDonald-Thompson test for pair-wise comparisons. Significance values are reported for $p < .05 (*)$, $p < .01 (**)$, and $p < .001 (***)$, respectively, abbreviated by the number of stars in parenthesis.

\begin{figure}
	\centering
	\includegraphics[width=\columnwidth]{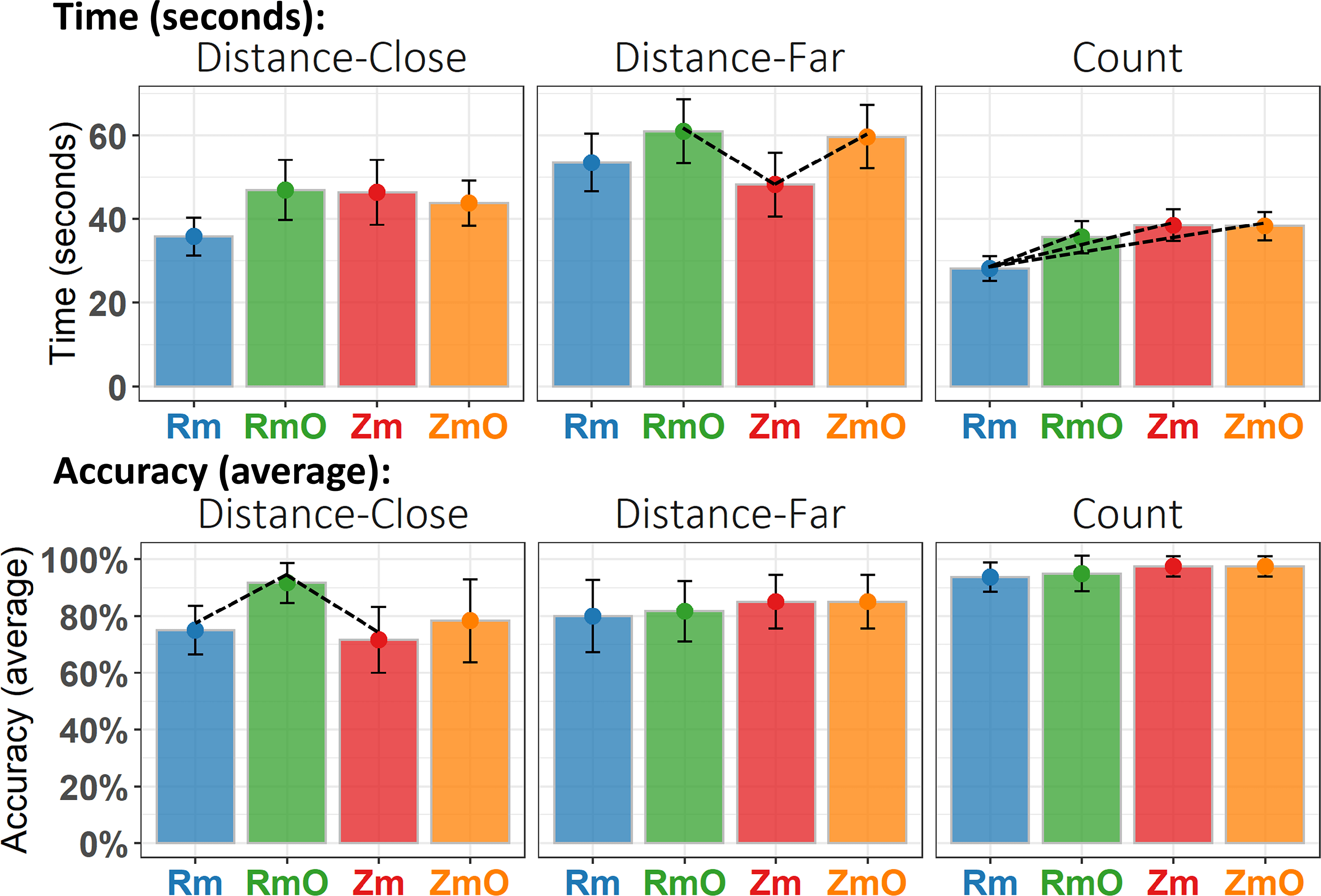}
	\caption{Results for \emph{time} (seconds) and accuracy by task. Confidence intervals indicate 95\% confidence for mean values. A dashed line indicates statistical significance for $p<.05$. }
	\label{fig:result-performance}
\end{figure}  
\begin{figure}
	\vspace{0.5em}
	\centering
	\includegraphics[width=\columnwidth]{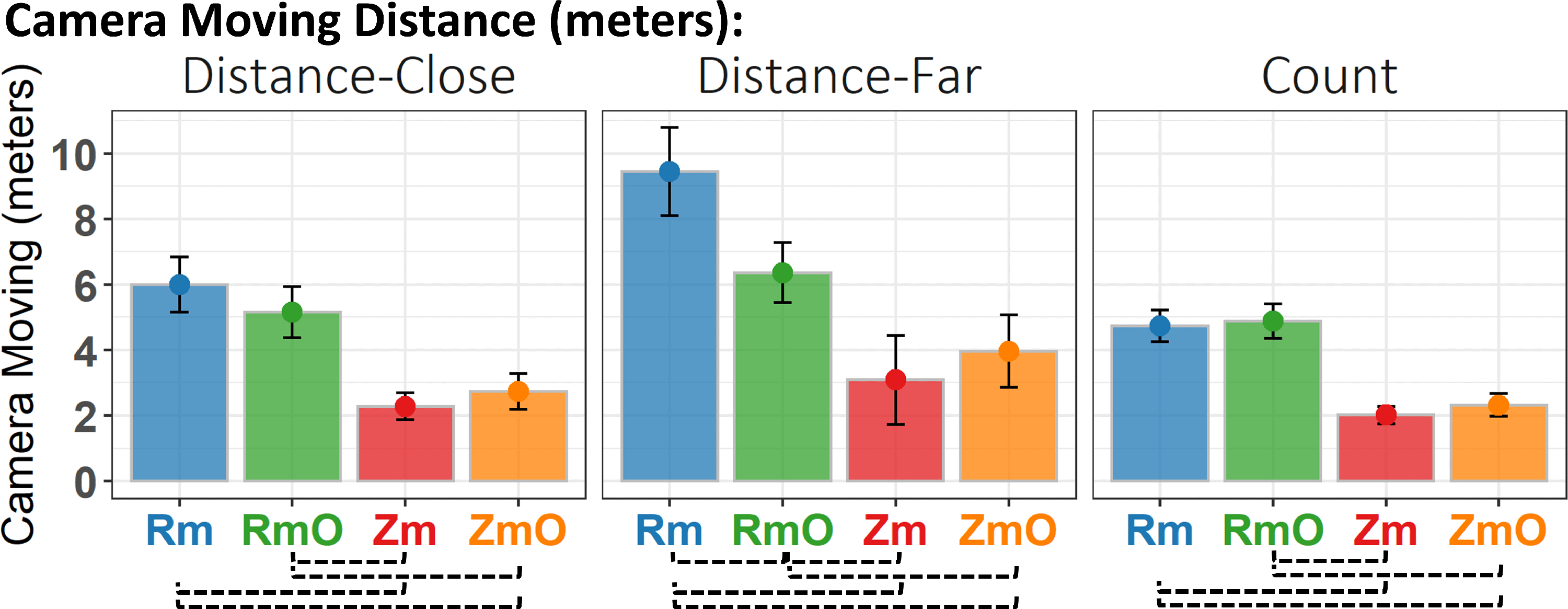}
	\caption{Camera movement distance per task trial. A dashed line indicates statistical significance for $p<.05$. }
	\label{fig:interaction-movement}
\end{figure}

\begin{figure*}
	\centering
	\includegraphics[width=\textwidth]{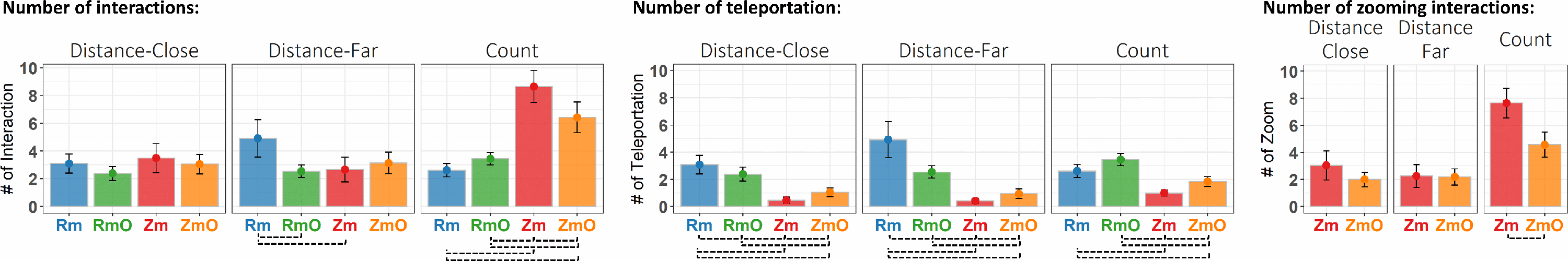}
	\caption{Number of interactions per task trial. The number of interactions is considered to be the sum of number of teleportations and number of zoomings. A dashed line indicates statistical significance for $p<.05$. }
	\label{fig:interaction-number}
\end{figure*} 

\begin{figure}
	\centering
	\includegraphics[width=\columnwidth]{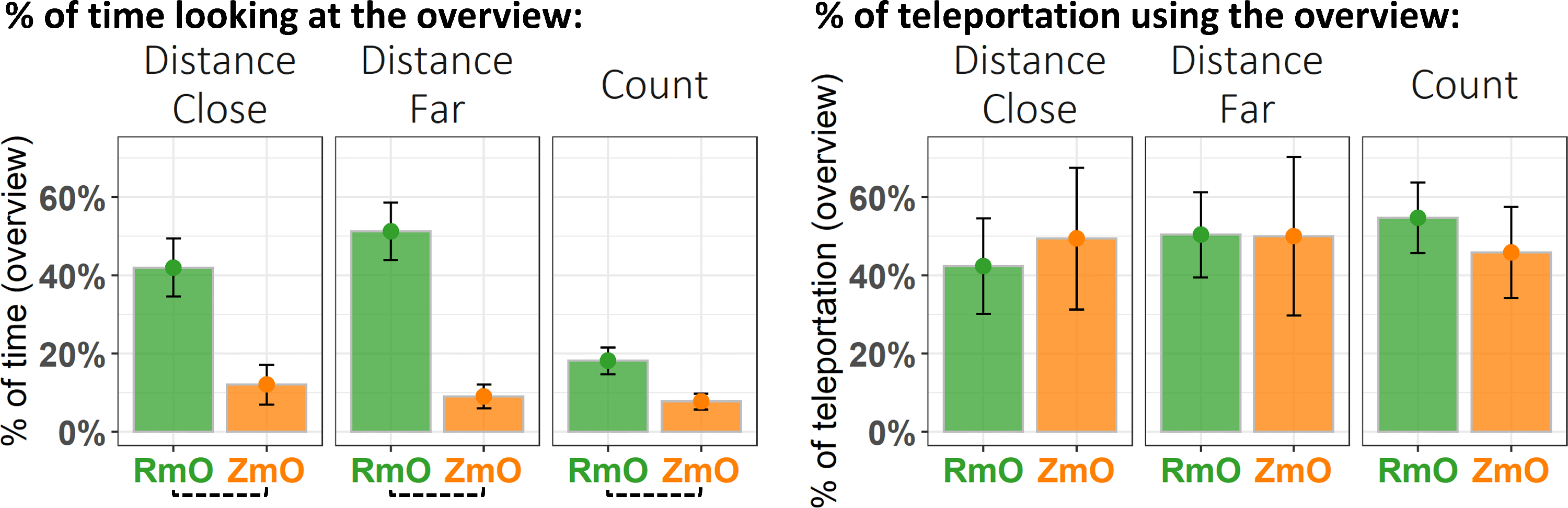}
	\caption{Average usage of the overview per trial. A dashed line indicates statistical significance for $p<.05$. }
	\label{fig:overview-usage}
\end{figure} 

\section{Results}
\label{sec:studyresults}
In this section, we first summarize self-reported strategies, then provide a pairwise comparison of performance (task time and accuracy) with the different visualization conditions. Finally we discuss interactions, user preference, and qualitative feedback.

\subsection{How did participants complete the tasks?}
\label{sec:result-strategy}
We asked participants to describe their strategies after each task. We found participants' strategies were relatively consistent.

\vspace{0.5em}
\noindent\textbf{The distance task:}
In \Room{}, most participants (14 out of 20) stayed within the visualization space,  within which, four participants explicitly mentioned that they used the pointer to ``\textit{jump}'' to the center of a pair of points as well as the center of two pairs. 
Six other participants stated they teleported outside the visualization space to have a better overview first, and then teleported back to the visualization space. 

In \RoomOD{}, most participants (13 out of 20) mentioned they mainly used the overview to find points. Eight participants used the overview to estimate the distance first, and then confirmed the answer in the detail view. 11 participants reported that they used the overview to teleport. Four participants used the same strategy they used in \Room{} to teleport to the center of a pair of points as well as the center of two pairs.

In \Zoom{}, most participants (17 out of 20) tried to find points in the small-sized view, and then compared the distance with an enlarged view. Other participants completed tasks with only small-sized views. 

In \ZoomOD{}, most participants (18 out of 20) used the most popular strategy in \Zoom{}, i.e., using the small-sized view to find points, and comparing distances with an enlarged view. In which, seven participants mentioned that they sometimes used the overview to teleport when the visualization was enlarged. Two other participants kept the visualization small all the time to answer the questions.

\vspace{0.5em}
\noindent\textbf{The count task:}
In \Room{}, most participants (14 out of 20) mainly physically walked to the place of interest in the space while the other six participants reported they mainly used pointer teleportation.

In \RoomOD{}, most participants (17 out of 20) mainly used the overview to teleport. Two other participants mainly used pointer teleportation, and one participant mainly walked.

In \Zoom{}, most participants (16 out of 20) zoomed in and out to reach different groups. Three other participants first enlarged the visualization, then grabbed and moved the view to get to groups. One participant first zoomed in and then used the pointer to teleport to groups.

In \ZoomOD{}, most participants (14 out of 20) zoomed in and out to complete the task and stated that they did not use the overview much. Six other participants first enlarged the visualization, and then used the overview to teleport to groups.

\subsection{Is having an overview beneficial?}
\noindent\textbf{\Room{} vs. \RoomOD{}:}
We found \Room{} was faster in the Count task ($**$). 
\Room{} also tended to be faster in the other tasks, but the differences were not statistically significant. 
We also found \RoomOD{} was more accurate in the Distance-Close task ($*$), see Fig.~\ref{fig:result-performance}. 
We believe the improved accuracy may come from the fact that the overview provided a different perspective as well as a different scale for the participants to confirm their answers. 
Eight participants explicitly reported using the overview for the distance comparison (see Sec.~\ref{sec:result-strategy}). 
We also found participants felt more confident with \RoomOD{} than \Room{} in the Distance task ($**$, see Fig.~\ref{fig:result-ratings}) which aligned with their higher accuracy in \RoomOD{}.

\noindent\textbf{\Zoom{} vs. \ZoomOD{}:}
We found similar performance between \Zoom{} and \ZoomOD{}, except that \ZoomOD{} was slower than \Zoom{} in the Distance-Far task condition ($*$, see Fig.~\ref{fig:result-performance}).
The very similar performance may because participants only use the overview occasionally in \ZoomOD{}. 
Apart from the overview, \Zoom{} and \ZoomOD{} share the same view and interactions. 
The limited use of the overview in \ZoomOD{} can be confirmed from both the users' strategy (see Sec.~\ref{sec:result-strategy}), and the fact that participants only spent around 10\% on average of their time looking at the overview (see Fig.~\ref{fig:overview-usage}).
This finding aligned with the results from Nekrasovski et al.~\cite{nekrasovski_evaluation_2006} where they found that presence of an overview did not affect the performance of a 2D zoomable hierarchical visualization (rubber sheet).

\noindent\textbf{Summary:}
Overall, adding an overview can increase task accuracy in some tasks, but may also introduce extra time cost. 
We also found an overview seems to be unnecessary in \Zoom{}, and adding one may even be distracting or disturbing for difficult tasks (e.g., the Distance-Far task). Due to the very similar performance between \Zoom{} and \ZoomOD{}, we did not include \ZoomOD{} in the following pair-wise comparisons explicitly.

\subsection{Visualization manipulation vs. Move (\Zoom{} vs. \Room{})}
We found \Zoom{} was slower than \Room{} in the Count task ($***$). It tended to be slower in the Distance-Close task but not significantly. \Zoom{} and \Room{} had similar time performance in the Distance-Far task. 
Our results partially align with the results from Lages and Bowman~\cite{lages_move_2018}. They found that the performance of walking and object manipulation in VR was affected by the gaming experience of participants. In essence, participants without significant gaming experience performed better with physical movement, while 3D manipulation enabled higher performance for participants with gaming experience. 
Most of our participants reported having no significant gaming experience (17 out of 20 play less than 2 hours of games per week). 
However, due to the limited number of participants with gaming experience in our user study, we are unable to draw statistical conclusions about this effect.

\subsection{Zooming vs. Overview+Detail (\Zoom{} vs. \RoomOD{})}
We found \Zoom{} and \RoomOD{} had similar time performance in the easier tasks (i.e., the Distance-Close and Count tasks), and \Zoom{} was faster than \RoomOD{} in the difficult task (i.e., the Distance-Far tasks, $*$, see Fig.~\ref{fig:result-performance}). 
Previous user studies on 2D displays also found a similar time performance between these two types of interfaces (e.g.,~\cite{plumlee_zooming_2006,ronne2011sizing}) in simple navigation tasks.
We also found \RoomOD{} was more accurate in the Distance-Close task ($*$). This finding is partially aligned with the study by Plumlee and Ware~\cite{plumlee_zooming_2006} where they also found overview+detail increased the accuracy compared to a zooming interface on a 2D display. They suggested that the benefit may be due to reduced visual working memory load by having an extra view in the overview+detail interface.

\subsection{How did participants move and interact?} 
\label{sec:results-interaction}
We found that participants had significantly more camera movement in \Room{} and \RoomOD{} than in \Zoom{} and \ZoomOD{} (all $***$). \Room{} also required more camera movement than \RoomOD{} in the Distance-Far task ($***$) (see Fig.~\ref{fig:interaction-movement}).
\added{Motion parallax is a likely explanation. It is key to depth perception in immersive environments: a stronger cue than stereopsis, as well as being key to resolving occlusion. As the size of visualization increases, you have to move further to get the same motion parallax benefits.}

\begin{wrapfigure}{r}{0.5\columnwidth}
    \centering
    \vspace{-1em}
    \includegraphics[width=0.48\columnwidth]{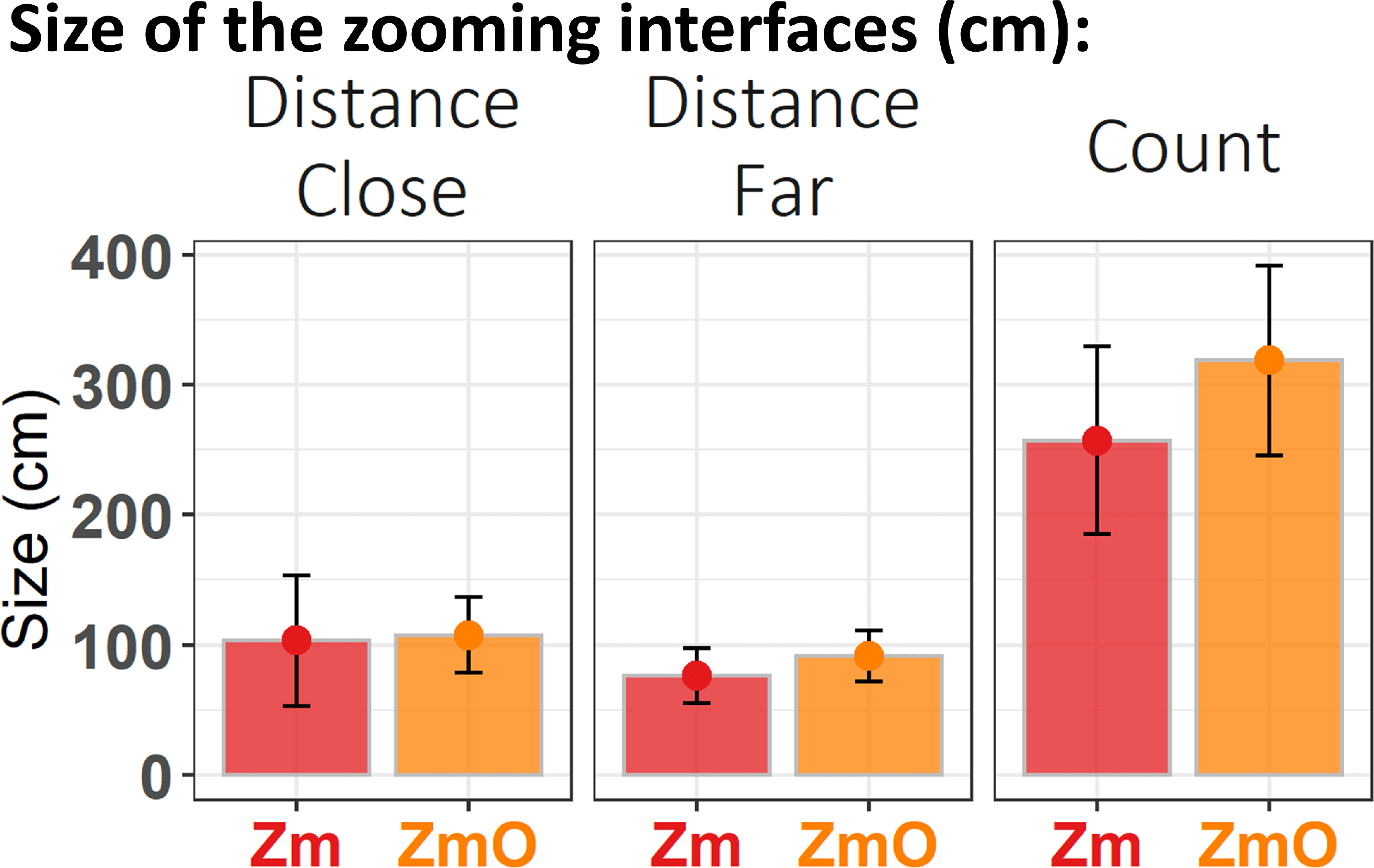}
    \caption{Size of the zoomable visualization per trial.}
    \label{fig:zoom-size}
    \vspace{0.5em}
\end{wrapfigure}

We also found participants teleported significantly more in \Room{} and \RoomOD{} than in \Zoom{} and \ZoomOD{} (all $***$). \Room{} also required more camera movement than \RoomOD{} in the Distance-Close and Distance-Far tasks ($***$). Participants also performed significantly more zooming interactions in \Zoom{} than \ZoomOD{} in the Count task ($***$). The size of the zooming interface was shown in Fig.~\ref{fig:zoom-size}.

We also added up the number of teleportation and zooming steps as the number of interactions. We found participants performed more interactions in \Room{} than in \RoomOD{} ($*$) and \Zoom{} ($***$) in the Distance-Far task. We also found \Zoom{} and \ZoomOD{} required more interactions than \Room{} and \RoomOD{} in the Count task (all $***$).

In summary, we found that \emph{overview} or \emph{zoom} reduced the number of required movements and teleportation compared to standard locomotion support alone. 
We also found \Zoom{} and \ZoomOD{} needed a significant amount of pinch-to-zoom interactions in the Count tasks.

\subsection{Which condition did participants prefer?} 
\label{sec:preference}
\begin{figure}
	\centering
	\includegraphics[width=0.65\columnwidth]{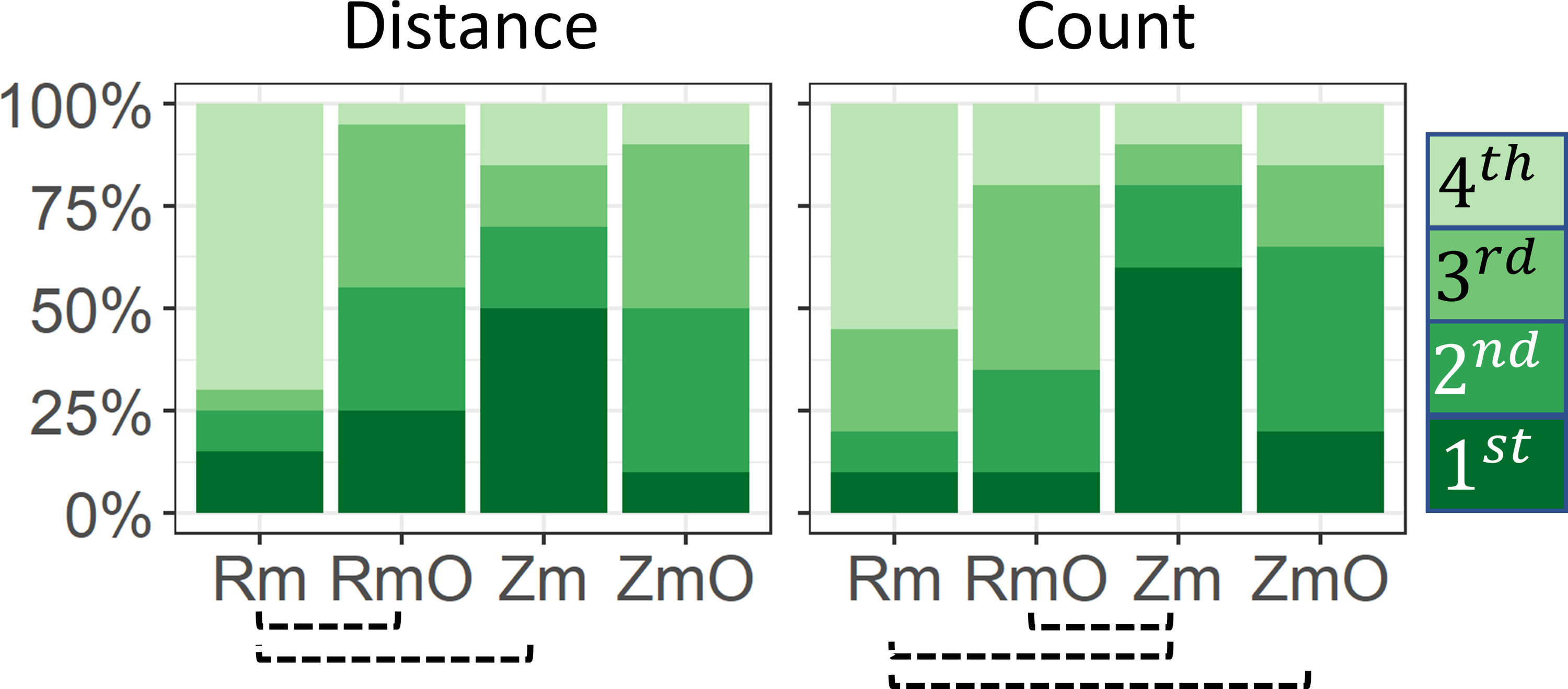}
	\caption{User preference ranking of each condition for two tasks (Distance and Count). Dashed lines indicate $p<.05$.}
	\label{fig:ranking}
\end{figure}

\begin{figure}
    \vspace{0.5em}
	\centering
	\includegraphics[width=0.55\columnwidth]{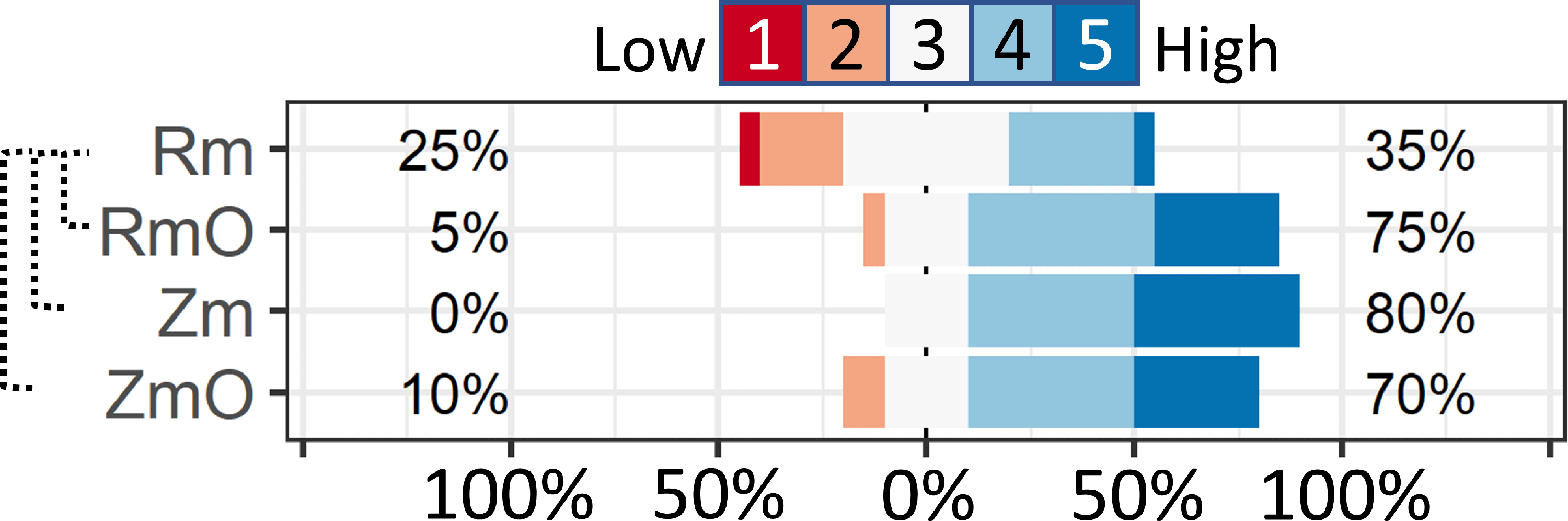}
	\caption{Overall usability ratings. Dashed lines indicate $p<.05$. 
	Percentage of negative and positive rankings is shown next to the bars.
	}
	\label{fig:usability}
\end{figure}

We asked participants to rank visualizations according to their preference for each task (see Fig.~\ref{fig:ranking}). 
For both the Distance and Count tasks, we found participants preferred \Zoom{} ($***$) and \RoomOD{} ($*$) over \Room{}. 
We also found \Zoom{} was preferred over \RoomOD{} ($*$).
Participants also rated the overall usability (see Fig.~\ref{fig:usability}). We found the \Room{} was considered to have lower usability than \RoomOD{} ($**$), \Zoom{} ($***$) and \ZoomOD{} ($*$).

\begin{figure}
	\centering
	\includegraphics[width=\columnwidth]{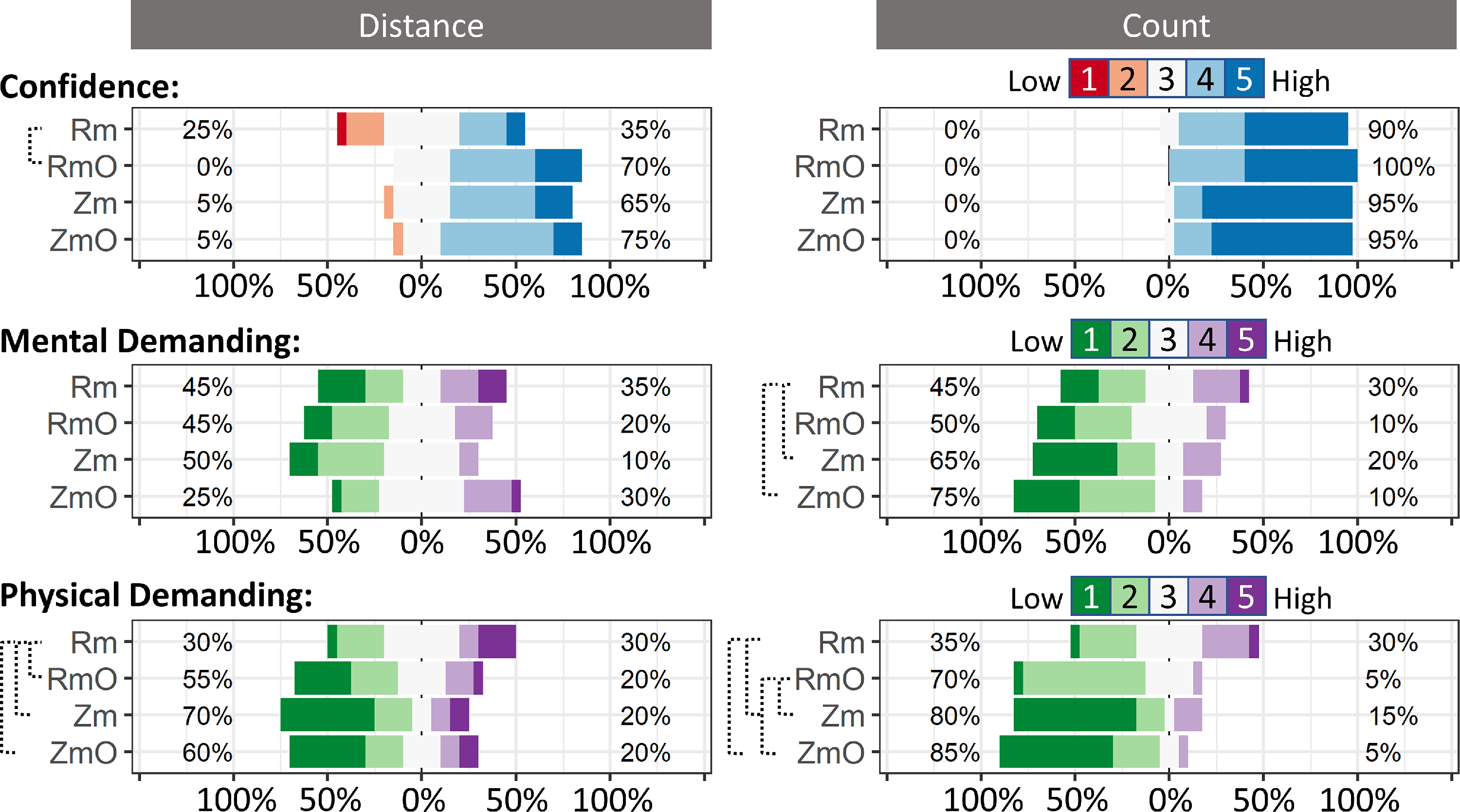}
	\caption{Confidence, mental demand, and physical demand ratings in a five-point-Likert scale for two tasks. Dashed lines indicate $p<.05$. 
	Percentage of negative and positive rankings is shown next to the bars.
	}
	\label{fig:result-ratings}
\end{figure}

\Zoom{} tended to be the most preferred visualization in our user study with more than 50\% of participants ranking it best in both tasks (see Fig.~\ref{fig:ranking}). \Zoom{} was also reported to be less demanding (see Fig.~\ref{fig:result-ratings}).
\Room{} was not preferred by our participants, even though it  generally performed well (see Fig.~\ref{fig:result-performance}). 
There could be two possible reasons: 
\emph{First}, participants felt \Room{} was more physical demanding (see Fig.~\ref{fig:result-ratings}) and the recorded movement data confirmed their subjective feeling (see Fig.~\ref{fig:interaction-movement}). 
\emph{Second}, with a fixed large scale single-view visualization, \Room{} was expected to have a high visual working memory load. The higher number of interactions and movements in \Room{} partially supports this assumption. 
Subjectively, participants also rated \Room{} to be more mentally demanding than \Zoom{} ($*$) and \ZoomOD{} ($*$) in the Count task.

\subsection{Qualitative user feedback}
\label{sec:feedback}
We asked participants to give feedback on the pros and cons of each design. 
\added{We clustered comments into groups for each visualization. In this subsection, we demonstrate representative ones along with the number of participants mentioned these similar comments.}

\Room{} was mentioned to be \emph{``close to real life''} by 11 participants. Among them, four participants explicitly reported it to be \emph{``immersive''}, three participants felt \emph{``more engaged''}, and three participants liked its fixed view: \emph{``it is easier to remember the points, whether in front or behind me.''}
However, ten participants also reported \emph{``it is difficult to find the points sometimes.''}

\RoomOD{} was considered an improvement over \Room{}.
15 participants reported that \emph{``the minimap was really useful for finding the points and moving around.''} Three participants also stated that \emph{``it is really good to have two scales [of views] at the same time.''}
However, two participants felt \emph{``overwhelmed''} by the interactions. Another two complained that \emph{``it breaks spatial continuity [when teleporting with the overview].''}

\Zoom{} was found to be \emph{``intuitive''} and \emph{``easy to use''} by 11 participants. Among them, three participants mentioned that they like the \emph{``flexibility''} of the interaction and feel they have \emph{``more control''}. One also commented that \emph{``it solves the problem of distance in a continuous way. Without losing the reference.''}
However, three commented it to be \emph{``not feeling real [comparing to \Room{} and \RoomOD{}].''} Two others  mentioned: \emph{``I may lose perspective after I move or zoom the view multiple times.''}

\ZoomOD{} was mainly compared to \Zoom{}. 
Six participants found \emph{``the minimap was good to jump around.''} 
However, 12 participants commented that \emph{``I did not use the minimap much.''} Five participants also complained that \emph{``it can be confusing as you have too many choices.''}


\subsection{Summary}
Of the four navigation methods that we tested, we found some significant differences in participant performance across the different tasks. However, there was no one navigation method that was best for every task.  The \emph{overview} increased accuracy for the \Room{} condition in the Distance-Close task, however, the \emph{overview} seemed to be an unnecessary distraction in the other tasks, and provided no benefit to the \Zoom{} condition. \Zoom{} was faster in the most difficult task (i.e., the Distance-Far task). Participants also clearly did not like the \Room{} condition.

\section{Discussion and Navigation Time-Cost Model}
\label{sec:timecostmodel}
Our study did not find a single best navigation method. In particular, analysis of the time data did not reveal a clear winner: for instance, \Zoom{} was faster in the Distance-Far task but slower in the Count task.
Furthermore, we found that \Room{} performed generally well in terms of time, but participants, for instance, complained about the difficulty of finding targets. This should have introduced extra time costs, but this was not reflected in the overall time.

To provide a more nuanced understanding of these mixed results, we now provide an initial exploratory analysis of the timing results based on previously suggested models of time cost for navigation~\cite{bowman_3d_2004,nilsson_natural_2018,plumlee_zooming_2002,plumlee_zooming_2006} and interactive visualization~\cite{lam_framework_2008,wang_baldonado_guidelines_2000}.
Navigation is a complex process with multiple components and the relevance of these components will vary in different tasks. By considering the cost of each component separately, we hope to better explain our results.
For example, while participants might spend more time identifying the targets with \Room{}, they may spend less time on other components, thereby compensating for this loss.
Based on the literature, the four most essential components in  models of navigation time-cost are:

\vspace{0.2em}
\noindent\textbf{Wayfinding} (term from~\cite{bowman_3d_2004,nilsson_natural_2018}): This is the process of finding the destination. Similar to \emph{decision costs to form goals} in~\cite{lam_framework_2008}.

\vspace{0.2em}
\noindent\textbf{Travel} (term from~\cite{bowman_3d_2004,nilsson_natural_2018}): This is the process of ``moving'' to the destination. It can be walking,  teleportation or manipulating the visualization to the desired form. Similar to \emph{physical-motion costs to execute sequences} in~\cite{lam_framework_2008} and \emph{transit between visits} in~\cite{plumlee_zooming_2002,plumlee_zooming_2006}

\vspace{0.2em}
\noindent\textbf{Number-of-travels} (term from~\cite{plumlee_zooming_2002,plumlee_zooming_2006}): Due to limited visual working memory~\cite{wang_baldonado_guidelines_2000}, completing a task can involve more than one travel to access information or to confirm the answer. 

\vspace{0.2em}
\noindent\textbf{Context-switching} (term from~\cite{wang_baldonado_guidelines_2000}): When the perceived view changes (either through physical movement or manipulating the visualization), the user must re-interpret it based on expectation. Similar to \emph{view-change costs to interpret perception} in~\cite{lam_framework_2008}.

In the rest of this section, we first analyze our visualization conditions with our navigation time-cost model. 
We then discuss the relative importance of the  components in our two tasks. 
Finally, we summarize our discussion by demonstrating how we can use our model to suggest visualization techniques for different tasks. We also demonstrate how we can use this model to identify the specific performance bottleneck of a visualization, and then propose potential strategies to improve it.

\subsection{Visualization Analysis}
\label{sec:vis-analysis}
\begin{figure}
	\centering
	\includegraphics[width=\columnwidth]{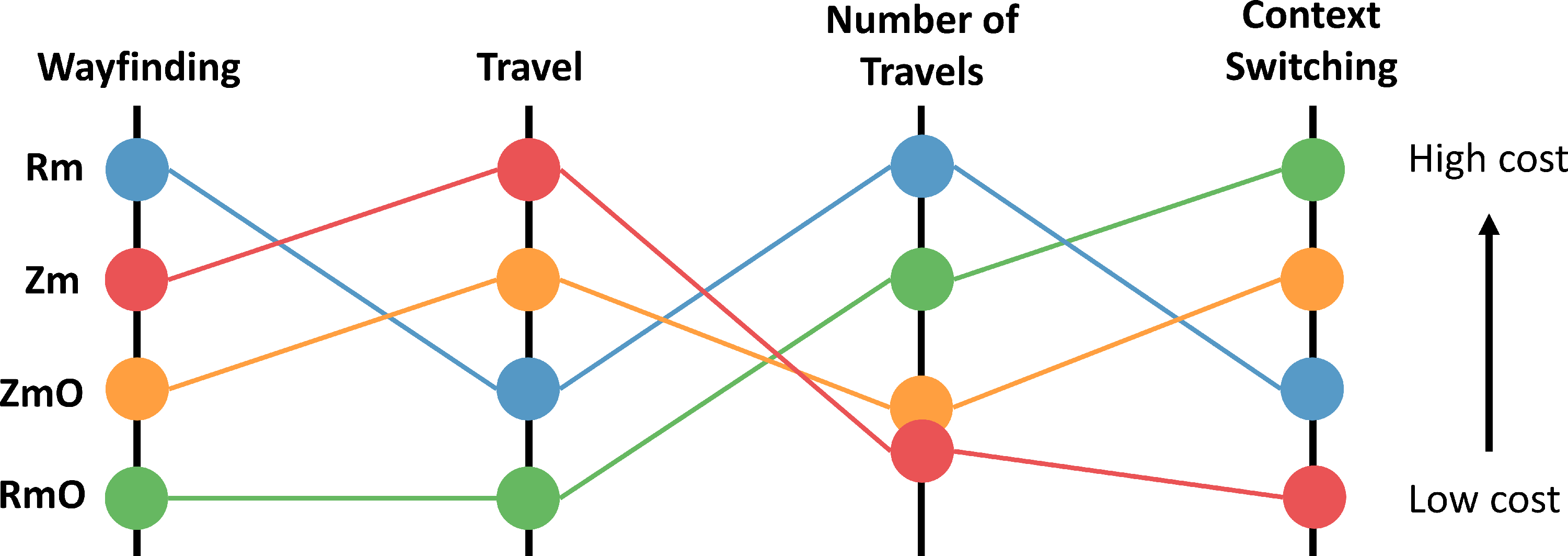}
	\caption{Navigation time-cost of our tested conditions broken down into four navigation components. 
	Positions are relative and qualitative, not based on precise metrics.}
	\label{fig:discussion-summary}
\end{figure}

In this subsection, we analyze the time cost (or performance) of our visualization conditions in terms of the above components.
The results of our analysis are demonstrated in Fig.~\ref{fig:discussion-summary}.

\vspace{0.3em}
\noindent\textbf{Wayfinding:}
The overview in \RoomOD{} could better facilitate the process of identifying targets compared to \Room{}. Participants' comments confirmed this, where 12 out of 20 mentioned \textit{``it is much easier to find the points [in \RoomOD{}] with the minimap [, rather than in \Room{}].''} For the same reason, we believe that the overview in \ZoomOD{} can better support identifying targets compared to \Zoom{}, especially when the visualization is enlarged. 
\Zoom{} supports searching targets in its small-sized state well. This is confirmed by the users' strategy where most participants located targets with size-reduced (or zoomed-out) \Zoom{} (see Sec.~\ref{sec:result-strategy}). However, participants lose the overview once they enlarge the visualization.
Participants clearly had difficulties in finding the targets in \Room{}(see Sec.~\ref{sec:feedback}).
In summary, we suggest that the time-cost of our tested visualizations in wayfinding is ordered as: \RoomOD{} $<$ \ZoomOD{} $<$ \Zoom{} $<$ \Room{}.

\vspace{0.3em}
\noindent\textbf{Travel:} 
We believe, in our relatively small-sized testing environment, physical movement in \Room{} might take less time than \Zoom{}. 
Familiarity with natural walking could make \Room{} outperform the relatively unnatural pinch-to-zoom gesture of \Zoom{} (i.e., people are unable to rescale a physical object in real life). 
This assumption was partially confirmed by previous studies (e.g., the ones reviewed by Ball et al.~\cite{ball_move_2007}) where physical movement gives better time performance using visualizations on large tiled displays. 
In \Room{}, apart from physical movement, we also provide pointer teleportation. However, pointer teleportation can only ease transit if the destination is within the users' FoV. In \RoomOD{}, the users can teleport to a place outside their FoV using the overview.
Compared to \Room{}, \RoomOD{} has a more flexible teleportation mechanism, which should result in a faster transit.
For the same reason, we believe \ZoomOD{} could outperform \Zoom{} in travel. 
In summary, we suggest that the time-cost of our tested visualizations in travel is ordered as: \RoomOD{} $<$ \Room{} $<$ \ZoomOD{} $<$ \Zoom{}.

\vspace{0.3em}
\noindent\textbf{Number-of-travels:} 
We use recorded interaction data as a proxy measure of the number-of-travels.
Overall, \Room{} clearly required significantly more physical movement and teleportations than other visualizations (see Sec.~\ref{sec:results-interaction}). 
\RoomOD{} also required more physical movement and teleportation than \Zoom{} (see Sec.~\ref{sec:results-interaction}). 
Although \Zoom{} required a large number of zooming in the Count task, we believe \RoomOD{} required an overall larger number of travels.
We found that the camera movement was similar in \Zoom{} and \ZoomOD{}. Meanwhile, \ZoomOD{} required more teleportation, while the number of performed zooming interactions was more in \Zoom{}. 
In summary, we suggest that our tested visualizations in number-of-travels is ordered as: \Zoom{} $\approx$ \ZoomOD{} $<$ \RoomOD{} $<$ \Room{}.

\vspace{0.3em}
\noindent\textbf{Context-switching:}
Physical movement is a spatial-continuous activity, which we consider to induce minimal context-switching costs. 
We also consider the ``pinch-to-zoom'' gesture a spatial-continuous traveling method which has similar performance compared to physical movement.  
Instant movement by teleportation introduces spatial disorientation and discontinuity~\cite{bakker_effects_2003,bowman_travel_1997}.
Furthermore, compared to the more predictable pointer teleportation where the destination is usually within the FoV, teleportation with the overview is expected to have a higher cost. 
Apart from teleportation, a user can also move based on the information in the overview (e.g., a user can identify the target is right of the current viewing direction in the overview and then turn right to find the target). This operation is also expected to be high-cost, as the user needs to visually link two separate display spaces. 
In \Room{}, participants performed more teleportations than in \Zoom{} (see Fig.~\ref{fig:interaction-number}), which we consider inducing greater context-switching costs. Participants also spent more time with the overview in \RoomOD{} than in \ZoomOD{} (see Fig.~\ref{fig:overview-usage}).
In summary, we suggest the time-cost of our tested visualizations in context-switching is ordered as: \Zoom{} $<$ \Room{} $<$ \ZoomOD{} $<$ \RoomOD{}.

\subsection{Task Analysis}
Based on the quantitative interaction data, qualitative feedback, and our observations in the user study, we infer the relative importance of the time-cost model components for our tested tasks (see Tab.~\ref{tab:weight-for-tasks}).

In the Count task, the targets are groups of points, so they are easy to find. Therefore, we believe that participants required minimal effort in wayfinding, and the number of travels is near identical across tested conditions. 
The context-switching effort should also be relatively low, as points in one group were all within the FoV. Thus, participants did not require frequent switching of views. Participants still needed to switch views when moving to the next target group, but the number of such switches is relatively low.

For the Distance task, there are only four colored targets, so the target points were more difficult to locate. Participants had to keep changing their viewing position and direction to find them. Moreover, the targets were mostly not within the participant's FoV at the same time, so participants had to switch views frequently to perform the comparison. The recorded interaction data shows that the Distance-Far task required more physical movement than the other two task conditions in Rm ($***$), RmO ($***$), and ZmO ($*$).

In summary, we suggest that the effort required for wayfinding, number-of-travels, and context-switching was higher in the Distance task than in the Count task. Within the Distance task, the Far condition requires more effort in these three components than the Close condition. Travel is an essential part in all navigation tasks, and we consider it to have a high weight in all conditions. 

\label{sec:task-analysis}
\begin{table}
	\centering
    \includegraphics[width=\columnwidth]{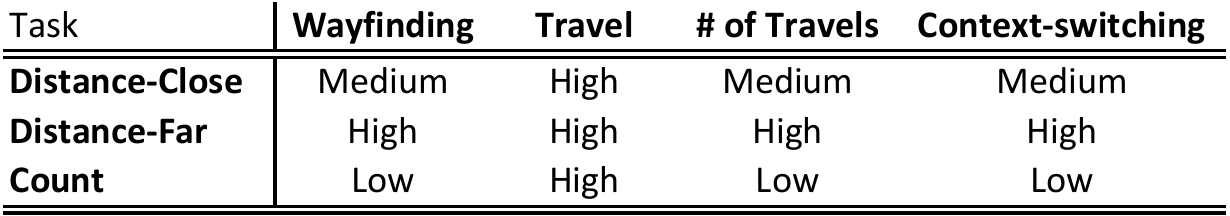}
    \caption{Relative importance of time-cost components for tasks.}
    \label{tab:weight-for-tasks}
    \vspace{-2em}
\end{table}

\subsection{Suggesting visualizations for tasks}
We demonstrate the potential of our navigation time-cost model to recommend visualization techniques for different tasks. We do this by explaining the overall time performance using the analysis results from Sec.~\ref{sec:vis-analysis} and~\ref{sec:task-analysis}.

For tasks that are less demanding on wayfinding and number-of-travels (e.g., the Distance-Close and Count tasks in our study), \Room{} is expected to have a good performance. 
This is because although \Room{} is not good at these two components, they have a limited amount of influence. On the other hand, \Room{} shows an overall good performance on the  more important component (i.e., travel).
For tasks that require a significant effort in number-of-travels and context-switch (e.g., the Distance-Far task in our study), \Zoom{} is a good choice, as it outperforms other conditions on these two components. 

\RoomOD{} has its advantages for wayfinding and travel, but the high cost in context-switching significantly influences its performance. For future studies, we should consider techniques that reduce the effort for context-switching in \RoomOD{}, e.g., animated  teleportation~\cite{bhandari_teleportation_2018,stoakley1995virtual}, 
or instead of always teleporting forwards, allowing users to interactively choose their viewing direction of the teleportation destination~\cite{funk_assessing_2019}.
We also propose a preliminary idea that when the user selects a target in the overview, a visual indicator will appear in the detail view to guide the user to the target. This is inspired by the work from Petford et al.~\cite{petford_comparison_2019}, which reviewed guiding techniques for out-of-view objects.

\section{Conclusion and Future Work}
\label{sec:conclusion}
We would recommend that developers of immersive visualization systems provide a variety of navigation methods to suit different tasks and environments.  For example, if the user has the capability to operate in a large open space, then there are definitely tasks (such as Distance Close) that will benefit from room-size navigation. However, in seated VR, the \emph{zoom} is going to be essential.  Our adaptation of the traditional \emph{overview} technique may be useful in room-size navigation for tasks that require operation at different scales, but such an overview should be easy to hide until required, to prevent distraction.

For future studies, a larger tracking space would support greater physical navigation but may also cause significant fatigue. Larger, more complex data may benefit the \emph{overview} more.
We also suggest that designs that can reduce context-switching cost in overview+detail interfaces are likely to improve its performance.
We also would like to design studies to  verify our navigation time-cost model systematically.

\acknowledgments{
This research was supported in part under KAUST Office of Sponsored Research (OSR) award OSR-2015-CCF-2533-01 and Australian Research Council’s Discovery Projects funding scheme DP180100755.
Yalong Yang was supported by a Harvard Physical Sciences and Engineering Accelerator Award. We also wish to thank all our participants for their time and our reviewers for their comments and feedback.
}

\bibliographystyle{abbrv-doi-hyperref}

\bibliography{reference}

\begin{thebibliography}{10}

\bibitem{abtahi_2019_Giant}
\href{https://doi.org/10.1145/3290605.3300752}{P.~Abtahi, M.~Gonzalez-Franco,
  E.~Ofek, and A.~Steed}.
\newblock \href{https://doi.org/10.1145/3290605.3300752}{I'm a {Giant}:
  {Walking} in {Large} {Virtual} {Environments} at {High} {Speed} {Gains}}.
\newblock \href{https://doi.org/10.1145/3290605.3300752}{In {\em Proceedings of
  the 2019 {CHI} {Conference} on {Human} {Factors} in {Computing} {Systems} -
  {CHI} '19}}, \href{https://doi.org/10.1145/3290605.3300752}{pp. 1--13}.
  \href{https://doi.org/10.1145/3290605.3300752}{ACM Press},
  \href{https://doi.org/10.1145/3290605.3300752}{Glasgow, Scotland Uk},
  \href{https://doi.org/10.1145/3290605.3300752}{2019}.
  \href{https://doi.org/10.1145/3290605.3300752}
{doi: {{%
10\hspace{.1pt}\discretionary{.}{%
}{.}\hspace{.4pt}1145\discretionary{/}{%
}{/}3290605\hspace{.1pt}\discretionary{.}{%
}{.}\hspace{.4pt}3300752}}}


\bibitem{bach_hologram_2018}
\href{https://doi.org/10.1109/TVCG.2017.2745941}{B.~Bach, R.~Sicat, J.~Beyer,
  M.~Cordeil, and H.~Pfister}.
\newblock \href{https://doi.org/10.1109/TVCG.2017.2745941}{The {Hologram} in
  {My} {Hand}: {How} {Effective} is {Interactive} {Exploration} of {3D}
  {Visualizations} in {Immersive} {Tangible} {Augmented} {Reality}?}
\newblock \href{https://doi.org/10.1109/TVCG.2017.2745941}{{\em IEEE
  Transactions on Visualization and Computer Graphics}},
  \href{https://doi.org/10.1109/TVCG.2017.2745941}{24(1):457--467},
  \href{https://doi.org/10.1109/TVCG.2017.2745941}{Jan. 2018}.
  \href{https://doi.org/10.1109/TVCG.2017.2745941}
{doi: {{%
10\hspace{.1pt}\discretionary{.}{%
}{.}\hspace{.4pt}1109\discretionary{/}{%
}{/}TVCG\hspace{.1pt}\discretionary{.}{%
}{.}\hspace{.4pt}2017\hspace{.1pt}\discretionary{.}{%
}{.}\hspace{.4pt}2745941}}}


\bibitem{bakker_effects_2003}
\href{https://doi.org/10.1518/hfes.45.1.160.27234}{N.~H. Bakker, P.~O.
  Passenier, and P.~J. Werkhoven}.
\newblock \href{https://doi.org/10.1518/hfes.45.1.160.27234}{Effects of
  {Head}-{Slaved} {Navigation} and the {Use} of {Teleports} on {Spatial}
  {Orientation} in {Virtual} {Environments}}.
\newblock \href{https://doi.org/10.1518/hfes.45.1.160.27234}{{\em Human
  Factors: The Journal of the Human Factors and Ergonomics Society}},
  \href{https://doi.org/10.1518/hfes.45.1.160.27234}{45(1):160--169},
  \href{https://doi.org/10.1518/hfes.45.1.160.27234}{Mar. 2003}.
  \href{https://doi.org/10.1518/hfes.45.1.160.27234}
{doi: {{%
10\hspace{.1pt}\discretionary{.}{%
}{.}\hspace{.4pt}1518\discretionary{/}{%
}{/}hfes\hspace{.1pt}\discretionary{.}{%
}{.}\hspace{.4pt}45\hspace{.1pt}\discretionary{.}{%
}{.}\hspace{.4pt}1\hspace{.1pt}\discretionary{.}{%
}{.}\hspace{.4pt}160\hspace{.1pt}\discretionary{.}{%
}{.}\hspace{.4pt}27234}}}


\bibitem{ball2005effects}
R.~Ball and C.~North.
\newblock Effects of tiled high-resolution display on basic visualization and
  navigation tasks.
\newblock In {\em CHI'05 extended abstracts on Human factors in computing
  systems}, pp. 1196--1199, 2005.

\bibitem{ball2008effects}
R.~Ball and C.~North.
\newblock The effects of peripheral vision and physical navigation on large
  scale visualization.
\newblock In {\em Proceedings of graphics interface 2008}, pp. 9--16, 2008.

\bibitem{ball_move_2007}
\href{https://doi.org/10.1145/1240624.1240656}{R.~Ball, C.~North, and D.~A.
  Bowman}.
\newblock \href{https://doi.org/10.1145/1240624.1240656}{Move to improve:
  promoting physical navigation to increase user performance with large
  displays}.
\newblock \href{https://doi.org/10.1145/1240624.1240656}{In {\em Proceedings of
  the {SIGCHI} {Conference} on {Human} {Factors} in {Computing} {Systems} -
  {CHI} '07}}, \href{https://doi.org/10.1145/1240624.1240656}{pp. 191--200}.
  \href{https://doi.org/10.1145/1240624.1240656}{ACM Press},
  \href{https://doi.org/10.1145/1240624.1240656}{San Jose, California, USA},
  \href{https://doi.org/10.1145/1240624.1240656}{2007}.
  \href{https://doi.org/10.1145/1240624.1240656}
{doi: {{%
10\hspace{.1pt}\discretionary{.}{%
}{.}\hspace{.4pt}1145\discretionary{/}{%
}{/}1240624\hspace{.1pt}\discretionary{.}{%
}{.}\hspace{.4pt}1240656}}}


\bibitem{batch2019there}
A.~Batch, A.~Cunningham, M.~Cordeil, N.~Elmqvist, T.~Dwyer, B.~H. Thomas, and
  K.~Marriott.
\newblock There is no spoon: Evaluating performance, space use, and presence
  with expert domain users in immersive analytics.
\newblock {\em IEEE transactions on visualization and computer graphics},
  26(1):536--546, 2019.

\bibitem{Bates2015}
\href{https://doi.org/10.18637/jss.v067.i01}{D.~Bates, M.~M\"{a}chler,
  B.~Bolker, and S.~Walker}.
\newblock \href{https://doi.org/10.18637/jss.v067.i01}{Fitting linear
  mixed-effects models using lme4}.
\newblock \href{https://doi.org/10.18637/jss.v067.i01}{{\em Journal of
  Statistical Software}}, \href{https://doi.org/10.18637/jss.v067.i01}{67(1)},
  \href{https://doi.org/10.18637/jss.v067.i01}{2015}.
  \href{https://doi.org/10.18637/jss.v067.i01}
{doi: {{%
10\hspace{.1pt}\discretionary{.}{%
}{.}\hspace{.4pt}18637\discretionary{/}{%
}{/}jss\hspace{.1pt}\discretionary{.}{%
}{.}\hspace{.4pt}v067\hspace{.1pt}\discretionary{.}{%
}{.}\hspace{.4pt}i01}}}


\bibitem{baudisch2002keeping}
P.~Baudisch, N.~Good, V.~Bellotti, and P.~Schraedley.
\newblock Keeping things in context: a comparative evaluation of focus plus
  context screens, overviews, and zooming.
\newblock In {\em Proceedings of the SIGCHI conference on Human factors in
  computing systems}, pp. 259--266, 2002.

\bibitem{bhandari_teleportation_2018}
\href{https://doi.org/10.20380/GI2018.22}{J.~Bhandari, P.~MacNeilage, and
  E.~Folmer}.
\newblock \href{https://doi.org/10.20380/GI2018.22}{Teleportation without
  {Spatial} {Disorientation} {Using} {Optical} {Flow} {Cues}}.
\newblock \href{https://doi.org/10.20380/GI2018.22}{In {\em Proceedings of the
  44th {Graphics} {Interface} {Conference}}},
  \href{https://doi.org/10.20380/GI2018.22}{{GI} '18},
  \href{https://doi.org/10.20380/GI2018.22}{pp. 162--167}.
  \href{https://doi.org/10.20380/GI2018.22}{Canadian Human-Computer
  Communications Society}, \href{https://doi.org/10.20380/GI2018.22}{Toronto,
  Canada}, \href{https://doi.org/10.20380/GI2018.22}{June 2018}.
  \href{https://doi.org/10.20380/GI2018.22}
{doi: {{%
10\hspace{.1pt}\discretionary{.}{%
}{.}\hspace{.4pt}20380\discretionary{/}{%
}{/}GI2018\hspace{.1pt}\discretionary{.}{%
}{.}\hspace{.4pt}22}}}


\bibitem{bowman_travel_1997}
\href{https://doi.org/10.1109/VRAIS.1997.583043}{D.~A. Bowman, D.~Koller, and
  L.~Hodges}.
\newblock \href{https://doi.org/10.1109/VRAIS.1997.583043}{Travel in immersive
  virtual environments: an evaluation of viewpoint motion control techniques}.
\newblock \href{https://doi.org/10.1109/VRAIS.1997.583043}{In {\em Proceedings
  of {IEEE} 1997 {Annual} {International} {Symposium} on {Virtual} {Reality}}},
  \href{https://doi.org/10.1109/VRAIS.1997.583043}{pp. 45--52,}.
  \href{https://doi.org/10.1109/VRAIS.1997.583043}{IEEE Comput. Soc. Press},
  \href{https://doi.org/10.1109/VRAIS.1997.583043}{Albuquerque, NM, USA},
  \href{https://doi.org/10.1109/VRAIS.1997.583043}{1997}.
  \href{https://doi.org/10.1109/VRAIS.1997.583043}
{doi: {{%
10\hspace{.1pt}\discretionary{.}{%
}{.}\hspace{.4pt}1109\discretionary{/}{%
}{/}VRAIS\hspace{.1pt}\discretionary{.}{%
}{.}\hspace{.4pt}1997\hspace{.1pt}\discretionary{.}{%
}{.}\hspace{.4pt}583043}}}


\bibitem{bowman1997travel}
D.~A. Bowman, D.~Koller, and L.~F. Hodges.
\newblock Travel in immersive virtual environments: An evaluation of viewpoint
  motion control techniques.
\newblock In {\em Proceedings of IEEE 1997 Annual International Symposium on
  Virtual Reality}, pp. 45--52. IEEE, 1997.

\bibitem{bowman_3d_2004}
D.~A. Bowman, E.~Kruijff, J.~J. LaViola, and I.~Poupyrev.
\newblock {\em {3D} {User} {Interfaces}: {Theory} and {Practice}}.
\newblock Addison Wesley Longman Publishing Co., Inc., USA, 2004.

\bibitem{burigat2013effectiveness}
S.~Burigat and L.~Chittaro.
\newblock On the effectiveness of overview+ detail visualization on mobile
  devices.
\newblock {\em Personal and ubiquitous computing}, 17(2):371--385, 2013.

\bibitem{burigat2008map}
S.~Burigat, L.~Chittaro, and E.~Parlato.
\newblock Map, diagram, and web page navigation on mobile devices: the
  effectiveness of zoomable user interfaces with overviews.
\newblock In {\em Proceedings of the 10th international conference on Human
  computer interaction with mobile devices and services}, pp. 147--156, 2008.

\bibitem{buring2006usability}
T.~B{\"u}ring, J.~Gerken, and H.~Reiterer.
\newblock Usability of overview-supported zooming on small screens with regard
  to individual differences in spatial ability.
\newblock In {\em Proceedings of the working conference on Advanced visual
  interfaces}, pp. 233--240, 2006.

\bibitem{card1999readings}
M.~Card.
\newblock {\em Readings in information visualization: using vision to think}.
\newblock Morgan Kaufmann, 1999.

\bibitem{chance1998locomotion}
S.~S. Chance, F.~Gaunet, A.~C. Beall, and J.~M. Loomis.
\newblock Locomotion mode affects the updating of objects encountered during
  travel: The contribution of vestibular and proprioceptive inputs to path
  integration.
\newblock {\em Presence}, 7(2):168--178, 1998.

\bibitem{cockburn2009review}
A.~Cockburn, A.~Karlson, and B.~B. Bederson.
\newblock A review of overview+ detail, zooming, and focus+ context interfaces.
\newblock {\em ACM Computing Surveys (CSUR)}, 41(1):1--31, 2009.

\bibitem{cockburn2004comparing}
A.~Cockburn and J.~Savage.
\newblock Comparing speed-dependent automatic zooming with traditional scroll,
  pan and zoom methods.
\newblock In {\em People and Computers XVII—Designing for Society}, pp.
  87--102. Springer, 2004.

\bibitem{cordeil_imaxes:_2017}
\href{https://doi.org/10.1145/3126594.3126613}{M.~Cordeil, A.~Cunningham,
  T.~Dwyer, B.~H. Thomas, and K.~Marriott}.
\newblock \href{https://doi.org/10.1145/3126594.3126613}{{ImAxes}: {Immersive}
  {Axes} as {Embodied} {Affordances} for {Interactive} {Multivariate} {Data}
  {Visualisation}}.
\newblock \href{https://doi.org/10.1145/3126594.3126613}{the 30th {Annual}
  {ACM} {Symposium} on {User} {Interface} {Software} and {Technology}},
  \href{https://doi.org/10.1145/3126594.3126613}{pp. 71--83},
  \href{https://doi.org/10.1145/3126594.3126613}{Oct. 2017}.
  \href{https://doi.org/10.1145/3126594.3126613}
{doi: {{%
10\hspace{.1pt}\discretionary{.}{%
}{.}\hspace{.4pt}1145\discretionary{/}{%
}{/}3126594\hspace{.1pt}\discretionary{.}{%
}{.}\hspace{.4pt}3126613}}}


\bibitem{dourish2001action}
P.~Dourish.
\newblock {\em Where the action is}.
\newblock MIT press Cambridge, 2001.

\bibitem{drogemuller_examining_2020}
\href{https://doi.org/10.1016/j.cola.2019.100937}{A.~Drogemuller,
  A.~Cunningham, J.~Walsh, B.~H. Thomas, M.~Cordeil, and W.~Ross}.
\newblock \href{https://doi.org/10.1016/j.cola.2019.100937}{Examining virtual
  reality navigation techniques for {3D} network visualisations}.
\newblock \href{https://doi.org/10.1016/j.cola.2019.100937}{{\em Journal of
  Computer Languages}},
  \href{https://doi.org/10.1016/j.cola.2019.100937}{56:100937},
  \href{https://doi.org/10.1016/j.cola.2019.100937}{Feb. 2020}.
  \href{https://doi.org/10.1016/j.cola.2019.100937}
{doi: {{%
10\hspace{.1pt}\discretionary{.}{%
}{.}\hspace{.4pt}1016\discretionary{/}{%
}{/}j\hspace{.1pt}\discretionary{.}{%
}{.}\hspace{.4pt}cola\hspace{.1pt}\discretionary{.}{%
}{.}\hspace{.4pt}2019\hspace{.1pt}\discretionary{.}{%
}{.}\hspace{.4pt}100937}}}


\bibitem{field2012discovering}
A.~Field, J.~Miles, and Z.~Field.
\newblock {\em Discovering statistics using R}.
\newblock Sage publications, 2012.

\bibitem{fonnet2018axes}
A.~Fonnet, T.~Vigier, Y.~Prie, G.~Cliquet, and F.~Picarougne.
\newblock Axes and coordinate systems representations for immersive analytics
  of multi-dimensional data.
\newblock In {\em 2018 International Symposium on Big Data Visual and Immersive
  Analytics (BDVA)}, pp. 1--10. IEEE, 2018.

\bibitem{fung2006treadmill}
J.~Fung, C.~L. Richards, F.~Malouin, B.~J. McFadyen, and A.~Lamontagne.
\newblock A treadmill and motion coupled virtual reality system for gait
  training post-stroke.
\newblock {\em CyberPsychology \& behavior}, 9(2):157--162, 2006.

\bibitem{funk_assessing_2019}
\href{https://doi.org/10.1145/3290605.3300377}{M.~Funk, F.~Müller,
  M.~Fendrich, M.~Shene, M.~Kolvenbach, N.~Dobbertin, S.~Günther, and
  M.~Mühlhäuser}.
\newblock \href{https://doi.org/10.1145/3290605.3300377}{Assessing the
  {Accuracy} of {Point} \& {Teleport} {Locomotion} with {Orientation}
  {Indication} for {Virtual} {Reality} using {Curved} {Trajectories}}.
\newblock \href{https://doi.org/10.1145/3290605.3300377}{In {\em Proceedings of
  the 2019 {CHI} {Conference} on {Human} {Factors} in {Computing} {Systems} -
  {CHI} '19}}, \href{https://doi.org/10.1145/3290605.3300377}{pp. 1--12}.
  \href{https://doi.org/10.1145/3290605.3300377}{ACM Press},
  \href{https://doi.org/10.1145/3290605.3300377}{Glasgow, Scotland Uk},
  \href{https://doi.org/10.1145/3290605.3300377}{2019}.
  \href{https://doi.org/10.1145/3290605.3300377}
{doi: {{%
10\hspace{.1pt}\discretionary{.}{%
}{.}\hspace{.4pt}1145\discretionary{/}{%
}{/}3290605\hspace{.1pt}\discretionary{.}{%
}{.}\hspace{.4pt}3300377}}}


\bibitem{hinckley_interaction_1998}
\href{https://doi.org/10.1145/288392.288572}{K.~Hinckley, M.~Czerwinski, and
  M.~Sinclair}.
\newblock \href{https://doi.org/10.1145/288392.288572}{Interaction and modeling
  techniques for desktop two-handed input}.
\newblock \href{https://doi.org/10.1145/288392.288572}{In {\em Proceedings of
  the 11th annual {ACM} symposium on {User} interface software and technology -
  {UIST} '98}}, \href{https://doi.org/10.1145/288392.288572}{pp. 49--58}.
  \href{https://doi.org/10.1145/288392.288572}{ACM Press},
  \href{https://doi.org/10.1145/288392.288572}{San Francisco, California,
  United States}, \href{https://doi.org/10.1145/288392.288572}{1998}.
  \href{https://doi.org/10.1145/288392.288572}
{doi: {{%
10\hspace{.1pt}\discretionary{.}{%
}{.}\hspace{.4pt}1145\discretionary{/}{%
}{/}288392\hspace{.1pt}\discretionary{.}{%
}{.}\hspace{.4pt}288572}}}


\bibitem{hornbaek2002navigation}
K.~Hornb{\ae}k, B.~B. Bederson, and C.~Plaisant.
\newblock Navigation patterns and usability of zoomable user interfaces with
  and without an overview.
\newblock {\em ACM Transactions on Computer-Human Interaction (TOCHI)},
  9(4):362--389, 2002.

\bibitem{hurter2018fiberclay}
C.~Hurter, N.~H. Riche, S.~M. Drucker, M.~Cordeil, R.~Alligier, and
  R.~Vuillemot.
\newblock Fiberclay: Sculpting three dimensional trajectories to reveal
  structural insights.
\newblock {\em IEEE Transactions on Visualization and Computer Graphics},
  25(1):704--714, 2018.

\bibitem{iwata2001gait}
H.~Iwata, H.~Yano, and F.~Nakaizumi.
\newblock Gait master: A versatile locomotion interface for uneven virtual
  terrain.
\newblock In {\em Proceedings IEEE Virtual Reality 2001}, pp. 131--137. IEEE,
  2001.

\bibitem{kraus_impact_2019}
\href{https://doi.org/10.1109/TVCG.2019.2934395}{M.~Kraus, N.~Weiler, D.~Oelke,
  J.~Kehrer, D.~A. Keim, and J.~Fuchs}.
\newblock \href{https://doi.org/10.1109/TVCG.2019.2934395}{The {Impact} of
  {Immersion} on {Cluster} {Identification} {Tasks}}.
\newblock \href{https://doi.org/10.1109/TVCG.2019.2934395}{{\em IEEE
  Transactions on Visualization and Computer Graphics}},
  \href{https://doi.org/10.1109/TVCG.2019.2934395}{pp. 1--1},
  \href{https://doi.org/10.1109/TVCG.2019.2934395}{2019}.
  \href{https://doi.org/10.1109/TVCG.2019.2934395}
{doi: {{%
10\hspace{.1pt}\discretionary{.}{%
}{.}\hspace{.4pt}1109\discretionary{/}{%
}{/}TVCG\hspace{.1pt}\discretionary{.}{%
}{.}\hspace{.4pt}2019\hspace{.1pt}\discretionary{.}{%
}{.}\hspace{.4pt}2934395}}}


\bibitem{kumar_browsing_1997}
\href{https://doi.org/10.1006/ijhc.1996.0085}{H.~P. Kumar, C.~Plaisant, and
  B.~Shneiderman}.
\newblock \href{https://doi.org/10.1006/ijhc.1996.0085}{Browsing hierarchical
  data with multi-level dynamic queries and pruning}.
\newblock \href{https://doi.org/10.1006/ijhc.1996.0085}{{\em International
  Journal of Human-Computer Studies}},
  \href{https://doi.org/10.1006/ijhc.1996.0085}{46(1):103--124},
  \href{https://doi.org/10.1006/ijhc.1996.0085}{Jan. 1997}.
  \href{https://doi.org/10.1006/ijhc.1996.0085}
{doi: {{%
10\hspace{.1pt}\discretionary{.}{%
}{.}\hspace{.4pt}1006\discretionary{/}{%
}{/}ijhc\hspace{.1pt}\discretionary{.}{%
}{.}\hspace{.4pt}1996\hspace{.1pt}\discretionary{.}{%
}{.}\hspace{.4pt}0085}}}


\bibitem{kwon_study_2016}
\href{https://doi.org/10.1109/TVCG.2016.2520921}{O.-H. Kwon, C.~Muelder,
  K.~Lee, and K.-L. Ma}.
\newblock \href{https://doi.org/10.1109/TVCG.2016.2520921}{A {Study} of
  {Layout}, {Rendering}, and {Interaction} {Methods} for {Immersive} {Graph}
  {Visualization}}.
\newblock \href{https://doi.org/10.1109/TVCG.2016.2520921}{{\em IEEE
  Transactions on Visualization and Computer Graphics}},
  \href{https://doi.org/10.1109/TVCG.2016.2520921}{22(7):1802--1815},
  \href{https://doi.org/10.1109/TVCG.2016.2520921}{July 2016}.
  \href{https://doi.org/10.1109/TVCG.2016.2520921}
{doi: {{%
10\hspace{.1pt}\discretionary{.}{%
}{.}\hspace{.4pt}1109\discretionary{/}{%
}{/}TVCG\hspace{.1pt}\discretionary{.}{%
}{.}\hspace{.4pt}2016\hspace{.1pt}\discretionary{.}{%
}{.}\hspace{.4pt}2520921}}}


\bibitem{lages_move_2018}
\href{https://doi.org/10.3389/fict.2018.00015}{W.~S. Lages and D.~A. Bowman}.
\newblock \href{https://doi.org/10.3389/fict.2018.00015}{Move the {Object} or
  {Move} {Myself}? {Walking} vs. {Manipulation} for the {Examination} of {3D}
  {Scientific} {Data}}.
\newblock \href{https://doi.org/10.3389/fict.2018.00015}{{\em Frontiers in
  ICT}}, \href{https://doi.org/10.3389/fict.2018.00015}{5:15},
  \href{https://doi.org/10.3389/fict.2018.00015}{July 2018}.
  \href{https://doi.org/10.3389/fict.2018.00015}
{doi: {{%
10\hspace{.1pt}\discretionary{.}{%
}{.}\hspace{.4pt}3389\discretionary{/}{%
}{/}fict\hspace{.1pt}\discretionary{.}{%
}{.}\hspace{.4pt}2018\hspace{.1pt}\discretionary{.}{%
}{.}\hspace{.4pt}00015}}}


\bibitem{lam_framework_2008}
\href{https://doi.org/10.1109/TVCG.2008.109}{H.~Lam}.
\newblock \href{https://doi.org/10.1109/TVCG.2008.109}{A {Framework} of
  {Interaction} {Costs} in {Information} {Visualization}}.
\newblock \href{https://doi.org/10.1109/TVCG.2008.109}{{\em IEEE Transactions
  on Visualization and Computer Graphics}},
  \href{https://doi.org/10.1109/TVCG.2008.109}{14(6):1149--1156},
  \href{https://doi.org/10.1109/TVCG.2008.109}{Nov. 2008}.
  \href{https://doi.org/10.1109/TVCG.2008.109}
{doi: {{%
10\hspace{.1pt}\discretionary{.}{%
}{.}\hspace{.4pt}1109\discretionary{/}{%
}{/}TVCG\hspace{.1pt}\discretionary{.}{%
}{.}\hspace{.4pt}2008\hspace{.1pt}\discretionary{.}{%
}{.}\hspace{.4pt}109}}}


\bibitem{laviola20173d}
J.~J. LaViola~Jr, E.~Kruijff, R.~P. McMahan, D.~Bowman, and I.~P. Poupyrev.
\newblock {\em 3D user interfaces: theory and practice}.
\newblock Addison-Wesley Professional, 2017.

\bibitem{lecun_gradient-based_1998}
\href{https://doi.org/10.1109/5.726791}{Y.~Lecun, L.~Bottou, Y.~Bengio, and
  P.~Haffner}.
\newblock \href{https://doi.org/10.1109/5.726791}{Gradient-based learning
  applied to document recognition}.
\newblock \href{https://doi.org/10.1109/5.726791}{{\em Proceedings of the
  IEEE}}, \href{https://doi.org/10.1109/5.726791}{86(11):2278--2324},
  \href{https://doi.org/10.1109/5.726791}{Nov. 1998}.
  \href{https://doi.org/10.1109/5.726791}
{doi: {{%
10\hspace{.1pt}\discretionary{.}{%
}{.}\hspace{.4pt}1109\discretionary{/}{%
}{/}5\hspace{.1pt}\discretionary{.}{%
}{.}\hspace{.4pt}726791}}}


\bibitem{Lenth2016}
\href{https://doi.org/10.18637/jss.v069.i01}{R.~V. Lenth}.
\newblock \href{https://doi.org/10.18637/jss.v069.i01}{Least-squares means:
  {The R Package lsmeans}}.
\newblock \href{https://doi.org/10.18637/jss.v069.i01}{{\em Journal of
  Statistical Software}}, \href{https://doi.org/10.18637/jss.v069.i01}{69(1)},
  \href{https://doi.org/10.18637/jss.v069.i01}{2016}.
  \href{https://doi.org/10.18637/jss.v069.i01}
{doi: {{%
10\hspace{.1pt}\discretionary{.}{%
}{.}\hspace{.4pt}18637\discretionary{/}{%
}{/}jss\hspace{.1pt}\discretionary{.}{%
}{.}\hspace{.4pt}v069\hspace{.1pt}\discretionary{.}{%
}{.}\hspace{.4pt}i01}}}


\bibitem{maaten_visualizing_2008}
\href{http://www.jmlr.org/papers/v9/vandermaaten08a.html}{L.~v.~d. Maaten and
  G.~Hinton}.
\newblock \href{http://www.jmlr.org/papers/v9/vandermaaten08a.html}{Visualizing
  {Data} using t-{SNE}}.
\newblock \href{http://www.jmlr.org/papers/v9/vandermaaten08a.html}{{\em
  Journal of Machine Learning Research}},
  \href{http://www.jmlr.org/papers/v9/vandermaaten08a.html}{9(Nov):2579--2605},
  \href{http://www.jmlr.org/papers/v9/vandermaaten08a.html}{2008}.

\bibitem{mackinlay1991perspective}
J.~D. Mackinlay, G.~G. Robertson, and S.~K. Card.
\newblock The perspective wall: Detail and context smoothly integrated.
\newblock In {\em Proceedings of the SIGCHI conference on Human factors in
  computing systems}, pp. 173--176, 1991.

\bibitem{mcgill_i_2017}
\href{https://doi.org/10.1145/3025453.3026046}{M.~McGill, A.~Ng, and
  S.~Brewster}.
\newblock \href{https://doi.org/10.1145/3025453.3026046}{I {Am} {The}
  {Passenger}: {How} {Visual} {Motion} {Cues} {Can} {Influence} {Sickness}
  {For} {In}-{Car} {VR}}.
\newblock \href{https://doi.org/10.1145/3025453.3026046}{In {\em Proceedings of
  the 2017 {CHI} {Conference} on {Human} {Factors} in {Computing} {Systems} -
  {CHI} '17}}, \href{https://doi.org/10.1145/3025453.3026046}{pp. 5655--5668}.
  \href{https://doi.org/10.1145/3025453.3026046}{ACM Press},
  \href{https://doi.org/10.1145/3025453.3026046}{Denver, Colorado, USA},
  \href{https://doi.org/10.1145/3025453.3026046}{2017}.
  \href{https://doi.org/10.1145/3025453.3026046}
{doi: {{%
10\hspace{.1pt}\discretionary{.}{%
}{.}\hspace{.4pt}1145\discretionary{/}{%
}{/}3025453\hspace{.1pt}\discretionary{.}{%
}{.}\hspace{.4pt}3026046}}}


\bibitem{mine1995isaac}
M.~R. Mine.
\newblock Isaac: A virtual environment tool for the interactive construction of
  virtual worlds.
\newblock Technical report, USA, 1995.

\bibitem{mine1995virtual}
M.~R. Mine.
\newblock Virtual environment interaction techniques.
\newblock {\em UNC Chapel Hill CS Dept}, 1995.

\bibitem{mine_moving_1997}
\href{https://doi.org/10.1145/258734.258747}{M.~R. Mine, F.~P. Brooks, and
  C.~H. Sequin}.
\newblock \href{https://doi.org/10.1145/258734.258747}{Moving objects in space:
  exploiting proprioception in virtual-environment interaction}.
\newblock \href{https://doi.org/10.1145/258734.258747}{In {\em Proceedings of
  the 24th annual conference on {Computer} graphics and interactive techniques
  - {SIGGRAPH} '97}}, \href{https://doi.org/10.1145/258734.258747}{pp. 19--26}.
  \href{https://doi.org/10.1145/258734.258747}{ACM Press},
  \href{https://doi.org/10.1145/258734.258747}{1997}.
  \href{https://doi.org/10.1145/258734.258747}
{doi: {{%
10\hspace{.1pt}\discretionary{.}{%
}{.}\hspace{.4pt}1145\discretionary{/}{%
}{/}258734\hspace{.1pt}\discretionary{.}{%
}{.}\hspace{.4pt}258747}}}


\bibitem{nam_worlds--wedges_2019}
\href{https://doi.org/10.1109/VR.2019.8797871}{J.~W. Nam, K.~McCullough,
  J.~Tveite, M.~M. Espinosa, C.~H. Perry, B.~T. Wilson, and D.~F. Keefe}.
\newblock \href{https://doi.org/10.1109/VR.2019.8797871}{Worlds-in-{Wedges}:
  {Combining} {Worlds}-in-{Miniature} and {Portals} to {Support} {Comparative}
  {Immersive} {Visualization} of {Forestry} {Data}}.
\newblock \href{https://doi.org/10.1109/VR.2019.8797871}{In {\em 2019 {IEEE}
  {Conference} on {Virtual} {Reality} and {3D} {User} {Interfaces} ({VR})}},
  \href{https://doi.org/10.1109/VR.2019.8797871}{pp. 747--755}.
  \href{https://doi.org/10.1109/VR.2019.8797871}{IEEE},
  \href{https://doi.org/10.1109/VR.2019.8797871}{Osaka, Japan},
  \href{https://doi.org/10.1109/VR.2019.8797871}{Mar. 2019}.
  \href{https://doi.org/10.1109/VR.2019.8797871}
{doi: {{%
10\hspace{.1pt}\discretionary{.}{%
}{.}\hspace{.4pt}1109\discretionary{/}{%
}{/}VR\hspace{.1pt}\discretionary{.}{%
}{.}\hspace{.4pt}2019\hspace{.1pt}\discretionary{.}{%
}{.}\hspace{.4pt}8797871}}}


\bibitem{nekrasovski_evaluation_2006}
\href{https://doi.org/10.1145/1124772.1124775}{D.~Nekrasovski, A.~Bodnar,
  J.~McGrenere, F.~Guimbretière, and T.~Munzner}.
\newblock \href{https://doi.org/10.1145/1124772.1124775}{An evaluation of pan
  \& zoom and rubber sheet navigation with and without an overview}.
\newblock \href{https://doi.org/10.1145/1124772.1124775}{In {\em Proceedings of
  the {SIGCHI} conference on {Human} {Factors} in computing systems - {CHI}
  '06}}, \href{https://doi.org/10.1145/1124772.1124775}{pp. 11--20}.
  \href{https://doi.org/10.1145/1124772.1124775}{ACM Press},
  \href{https://doi.org/10.1145/1124772.1124775}{Montreal, Québec, Canada},
  \href{https://doi.org/10.1145/1124772.1124775}{2006}.
  \href{https://doi.org/10.1145/1124772.1124775}
{doi: {{%
10\hspace{.1pt}\discretionary{.}{%
}{.}\hspace{.4pt}1145\discretionary{/}{%
}{/}1124772\hspace{.1pt}\discretionary{.}{%
}{.}\hspace{.4pt}1124775}}}


\bibitem{nilsson_natural_2018}
\href{https://doi.org/10.1145/3180658}{N.~C. Nilsson, S.~Serafin, F.~Steinicke,
  and R.~Nordahl}.
\newblock \href{https://doi.org/10.1145/3180658}{Natural {Walking} in {Virtual}
  {Reality}: {A} {Review}}.
\newblock \href{https://doi.org/10.1145/3180658}{{\em Computers in
  Entertainment}}, \href{https://doi.org/10.1145/3180658}{16(2):1--22},
  \href{https://doi.org/10.1145/3180658}{Apr. 2018}.
  \href{https://doi.org/10.1145/3180658}
{doi: {{%
10\hspace{.1pt}\discretionary{.}{%
}{.}\hspace{.4pt}1145\discretionary{/}{%
}{/}3180658}}}


\bibitem{petford_comparison_2019}
\href{https://doi.org/10.1145/3290605.3300288}{J.~Petford, I.~Carson, M.~A.
  Nacenta, and C.~Gutwin}.
\newblock \href{https://doi.org/10.1145/3290605.3300288}{A {Comparison} of
  {Notification} {Techniques} for {Out}-of-{View} {Objects} in
  {Full}-{Coverage} {Displays}}.
\newblock \href{https://doi.org/10.1145/3290605.3300288}{In {\em Proceedings of
  the 2019 {CHI} {Conference} on {Human} {Factors} in {Computing} {Systems} -
  {CHI} '19}}, \href{https://doi.org/10.1145/3290605.3300288}{pp. 1--13}.
  \href{https://doi.org/10.1145/3290605.3300288}{ACM Press},
  \href{https://doi.org/10.1145/3290605.3300288}{Glasgow, Scotland Uk},
  \href{https://doi.org/10.1145/3290605.3300288}{2019}.
  \href{https://doi.org/10.1145/3290605.3300288}
{doi: {{%
10\hspace{.1pt}\discretionary{.}{%
}{.}\hspace{.4pt}1145\discretionary{/}{%
}{/}3290605\hspace{.1pt}\discretionary{.}{%
}{.}\hspace{.4pt}3300288}}}


\bibitem{plumlee_zooming_2002}
\href{https://doi.org/10.1145/1556262.1556270}{M.~Plumlee and C.~Ware}.
\newblock \href{https://doi.org/10.1145/1556262.1556270}{Zooming, multiple
  windows, and visual working memory}.
\newblock \href{https://doi.org/10.1145/1556262.1556270}{In {\em Proceedings of
  the {Working} {Conference} on {Advanced} {Visual} {Interfaces} - {AVI} '02}},
  \href{https://doi.org/10.1145/1556262.1556270}{p.~59}.
  \href{https://doi.org/10.1145/1556262.1556270}{ACM Press},
  \href{https://doi.org/10.1145/1556262.1556270}{Trento, Italy},
  \href{https://doi.org/10.1145/1556262.1556270}{2002}.
  \href{https://doi.org/10.1145/1556262.1556270}
{doi: {{%
10\hspace{.1pt}\discretionary{.}{%
}{.}\hspace{.4pt}1145\discretionary{/}{%
}{/}1556262\hspace{.1pt}\discretionary{.}{%
}{.}\hspace{.4pt}1556270}}}


\bibitem{plumlee_zooming_2006}
\href{https://doi.org/10.1145/1165734.1165736}{M.~D. Plumlee and C.~Ware}.
\newblock \href{https://doi.org/10.1145/1165734.1165736}{Zooming versus
  multiple window interfaces: {Cognitive} costs of visual comparisons}.
\newblock \href{https://doi.org/10.1145/1165734.1165736}{{\em ACM Transactions
  on Computer-Human Interaction (TOCHI)}},
  \href{https://doi.org/10.1145/1165734.1165736}{13(2):179--209},
  \href{https://doi.org/10.1145/1165734.1165736}{June 2006}.
  \href{https://doi.org/10.1145/1165734.1165736}
{doi: {{%
10\hspace{.1pt}\discretionary{.}{%
}{.}\hspace{.4pt}1145\discretionary{/}{%
}{/}1165734\hspace{.1pt}\discretionary{.}{%
}{.}\hspace{.4pt}1165736}}}


\bibitem{prouzeau_scaptics_2019}
\href{https://doi.org/10.1145/3290605.3300555}{A.~Prouzeau, M.~Cordeil,
  C.~Robin, B.~Ens, B.~H. Thomas, and T.~Dwyer}.
\newblock \href{https://doi.org/10.1145/3290605.3300555}{Scaptics and
  {Highlight}-{Planes}: {Immersive} {Interaction} {Techniques} for {Finding}
  {Occluded} {Features} in {3D} {Scatterplots}}.
\newblock \href{https://doi.org/10.1145/3290605.3300555}{In {\em Proceedings of
  the 2019 {CHI} {Conference} on {Human} {Factors} in {Computing} {Systems} -
  {CHI} '19}}, \href{https://doi.org/10.1145/3290605.3300555}{pp. 1--12}.
  \href{https://doi.org/10.1145/3290605.3300555}{ACM Press},
  \href{https://doi.org/10.1145/3290605.3300555}{Glasgow, Scotland Uk},
  \href{https://doi.org/10.1145/3290605.3300555}{2019}.
  \href{https://doi.org/10.1145/3290605.3300555}
{doi: {{%
10\hspace{.1pt}\discretionary{.}{%
}{.}\hspace{.4pt}1145\discretionary{/}{%
}{/}3290605\hspace{.1pt}\discretionary{.}{%
}{.}\hspace{.4pt}3300555}}}


\bibitem{radle2013effect}
R.~R{\"a}dle, H.-C. Jetter, S.~Butscher, and H.~Reiterer.
\newblock The effect of egocentric body movements on users' navigation
  performance and spatial memory in zoomable user interfaces.
\newblock In {\em Proceedings of the 2013 ACM international conference on
  Interactive tabletops and surfaces}, pp. 23--32, 2013.

\bibitem{raja2004exploring}
D.~Raja, D.~Bowman, J.~Lucas, and C.~North.
\newblock Exploring the benefits of immersion in abstract information
  visualization.
\newblock In {\em Proc. Immersive Projection Technology Workshop}, pp. 61--69,
  2004.

\bibitem{ronne2011sizing}
M.~R{\o}nne~Jakobsen and K.~Hornb{\ae}k.
\newblock Sizing up visualizations: effects of display size in focus+ context,
  overview+ detail, and zooming interfaces.
\newblock In {\em Proceedings of the SIGCHI Conference on Human Factors in
  Computing Systems}, pp. 1451--1460, 2011.

\bibitem{ruddle2009benefits}
R.~A. Ruddle and S.~Lessels.
\newblock The benefits of using a walking interface to navigate virtual
  environments.
\newblock {\em ACM Transactions on Computer-Human Interaction (TOCHI)},
  16(1):1--18, 2009.

\bibitem{ruddle2011walking}
R.~A. Ruddle, E.~Volkova, and H.~H. B{\"u}lthoff.
\newblock Walking improves your cognitive map in environments that are
  large-scale and large in extent.
\newblock {\em ACM Transactions on Computer-Human Interaction (TOCHI)},
  18(2):1--20, 2011.

\bibitem{sarikaya_scatterplots_2018}
\href{https://doi.org/10.1109/TVCG.2017.2744184}{A.~Sarikaya and M.~Gleicher}.
\newblock \href{https://doi.org/10.1109/TVCG.2017.2744184}{Scatterplots:
  {Tasks}, {Data}, and {Designs}}.
\newblock \href{https://doi.org/10.1109/TVCG.2017.2744184}{{\em IEEE
  Transactions on Visualization and Computer Graphics}},
  \href{https://doi.org/10.1109/TVCG.2017.2744184}{24(1):402--412},
  \href{https://doi.org/10.1109/TVCG.2017.2744184}{Jan. 2018}.
  \href{https://doi.org/10.1109/TVCG.2017.2744184}
{doi: {{%
10\hspace{.1pt}\discretionary{.}{%
}{.}\hspace{.4pt}1109\discretionary{/}{%
}{/}TVCG\hspace{.1pt}\discretionary{.}{%
}{.}\hspace{.4pt}2017\hspace{.1pt}\discretionary{.}{%
}{.}\hspace{.4pt}2744184}}}


\bibitem{satriadi2019augmented}
K.~A. Satriadi, B.~Ens, M.~Cordeil, B.~Jenny, T.~Czauderna, and W.~Willett.
\newblock Augmented reality map navigation with freehand gestures.
\newblock In {\em 2019 IEEE Conference on Virtual Reality and 3D User
  Interfaces (VR)}, pp. 593--603. IEEE, 2019.

\bibitem{shneiderman1996eyes}
B.~Shneiderman.
\newblock The eyes have it: A task by data type taxonomy for information
  visualizations.
\newblock In {\em Proceedings 1996 IEEE symposium on visual languages}, pp.
  336--343. IEEE, 1996.

\bibitem{simpson2017take}
M.~Simpson, J.~Zhao, and A.~Klippel.
\newblock Take a walk: Evaluating movement types for data visualization in
  immersive virtual reality.
\newblock In {\em Workshop on Immersive Analytics, IEEE Vis}, 2017.

\bibitem{sorger2019immersive}
J.~Sorger, M.~Waldner, W.~Knecht, and A.~Arleo.
\newblock Immersive analytics of large dynamic networks via overview and detail
  navigation.
\newblock {\em arXiv preprint arXiv:1910.06825}, 2019.

\bibitem{stoakley1995virtual}
\href{https://doi.org/10.1145/223904.223938}{R.~Stoakley, M.~J. Conway, and
  R.~Pausch}.
\newblock \href{https://doi.org/10.1145/223904.223938}{Virtual reality on a
  {WIM}: interactive worlds in miniature}.
\newblock \href{https://doi.org/10.1145/223904.223938}{In {\em Proceedings of
  the {SIGCHI} conference on {Human} factors in computing systems - {CHI}
  '95}}, \href{https://doi.org/10.1145/223904.223938}{pp. 265--272}.
  \href{https://doi.org/10.1145/223904.223938}{ACM Press},
  \href{https://doi.org/10.1145/223904.223938}{Denver, Colorado, United
  States}, \href{https://doi.org/10.1145/223904.223938}{1995}.
  \href{https://doi.org/10.1145/223904.223938}
{doi: {{%
10\hspace{.1pt}\discretionary{.}{%
}{.}\hspace{.4pt}1145\discretionary{/}{%
}{/}223904\hspace{.1pt}\discretionary{.}{%
}{.}\hspace{.4pt}223938}}}


\bibitem{embedded-projector}
{TensorFlow, Google}.
\newblock {Embedding Projector}.
\newblock \url{https://projector.tensorflow.org/}, 2020.
\newblock Online; accessed March 2020.

\bibitem{trueba2009complexity}
R.~Trueba, C.~Andujar, and F.~Argelaguet.
\newblock Complexity and occlusion management for the world-in-miniature
  metaphor.
\newblock In {\em International Symposium on Smart Graphics}, pp. 155--166.
  Springer, 2009.

\bibitem{filho_virtualdesk:_2018}
\href{https://doi.org/10.1111/cgf.13430}{J.~A. Wagner~Filho, C.~Freitas, and
  L.~Nedel}.
\newblock \href{https://doi.org/10.1111/cgf.13430}{{VirtualDesk}: {A}
  {Comfortable} and {Efficient} {Immersive} {Information} {Visualization}
  {Approach}}.
\newblock \href{https://doi.org/10.1111/cgf.13430}{{\em Computer Graphics
  Forum}}, \href{https://doi.org/10.1111/cgf.13430}{37(3):415--426},
  \href{https://doi.org/10.1111/cgf.13430}{June 2018}.
  \href{https://doi.org/10.1111/cgf.13430}
{doi: {{%
10\hspace{.1pt}\discretionary{.}{%
}{.}\hspace{.4pt}1111\discretionary{/}{%
}{/}cgf\hspace{.1pt}\discretionary{.}{%
}{.}\hspace{.4pt}13430}}}


\bibitem{wagner_filho_immersive_2018}
\href{https://doi.org/10.1109/VR.2018.8447558}{J.~A. Wagner~Filho, M.~F. Rey,
  C.~M. D.~S. Freitas, and L.~Nedel}.
\newblock \href{https://doi.org/10.1109/VR.2018.8447558}{Immersive
  {Visualization} of {Abstract} {Information}: {An} {Evaluation} on
  {Dimensionally}-{Reduced} {Data} {Scatterplots}}.
\newblock \href{https://doi.org/10.1109/VR.2018.8447558}{In {\em 2018 {IEEE}
  {Conference} on {Virtual} {Reality} and {3D} {User} {Interfaces} ({VR})}},
  \href{https://doi.org/10.1109/VR.2018.8447558}{pp. 483--490}.
  \href{https://doi.org/10.1109/VR.2018.8447558}{IEEE},
  \href{https://doi.org/10.1109/VR.2018.8447558}{Mar. 2018}.
  \href{https://doi.org/10.1109/VR.2018.8447558}
{doi: {{%
10\hspace{.1pt}\discretionary{.}{%
}{.}\hspace{.4pt}1109\discretionary{/}{%
}{/}VR\hspace{.1pt}\discretionary{.}{%
}{.}\hspace{.4pt}2018\hspace{.1pt}\discretionary{.}{%
}{.}\hspace{.4pt}8447558}}}


\bibitem{wagner_filho_evaluating_2019}
\href{https://doi.org/10.1109/TVCG.2019.2934415}{J.~A. Wagner~Filho,
  W.~Stuerzlinger, and L.~Nedel}.
\newblock \href{https://doi.org/10.1109/TVCG.2019.2934415}{Evaluating an
  {Immersive} {Space}-{Time} {Cube} {Geovisualization} for {Intuitive}
  {Trajectory} {Data} {Exploration}}.
\newblock \href{https://doi.org/10.1109/TVCG.2019.2934415}{{\em IEEE
  Transactions on Visualization and Computer Graphics}},
  \href{https://doi.org/10.1109/TVCG.2019.2934415}{26(1):514--524},
  \href{https://doi.org/10.1109/TVCG.2019.2934415}{2019}.
  \href{https://doi.org/10.1109/TVCG.2019.2934415}
{doi: {{%
10\hspace{.1pt}\discretionary{.}{%
}{.}\hspace{.4pt}1109\discretionary{/}{%
}{/}TVCG\hspace{.1pt}\discretionary{.}{%
}{.}\hspace{.4pt}2019\hspace{.1pt}\discretionary{.}{%
}{.}\hspace{.4pt}2934415}}}


\bibitem{wang_baldonado_guidelines_2000}
\href{https://doi.org/10.1145/345513.345271}{M.~Q. Wang~Baldonado, A.~Woodruff,
  and A.~Kuchinsky}.
\newblock \href{https://doi.org/10.1145/345513.345271}{Guidelines for using
  multiple views in information visualization}.
\newblock \href{https://doi.org/10.1145/345513.345271}{In {\em Proceedings of
  the working conference on {Advanced} visual interfaces - {AVI} '00}},
  \href{https://doi.org/10.1145/345513.345271}{pp. 110--119}.
  \href{https://doi.org/10.1145/345513.345271}{ACM Press},
  \href{https://doi.org/10.1145/345513.345271}{Palermo, Italy},
  \href{https://doi.org/10.1145/345513.345271}{2000}.
  \href{https://doi.org/10.1145/345513.345271}
{doi: {{%
10\hspace{.1pt}\discretionary{.}{%
}{.}\hspace{.4pt}1145\discretionary{/}{%
}{/}345513\hspace{.1pt}\discretionary{.}{%
}{.}\hspace{.4pt}345271}}}


\bibitem{wei_evaluating_2020}
\href{https://doi.org/10.1109/TVCG.2019.2934208}{Y.~Wei, H.~Mei, Y.~Zhao,
  S.~Zhou, B.~Lin, H.~Jiang, and W.~Chen}.
\newblock \href{https://doi.org/10.1109/TVCG.2019.2934208}{Evaluating
  {Perceptual} {Bias} {During} {Geometric} {Scaling} of {Scatterplots}}.
\newblock \href{https://doi.org/10.1109/TVCG.2019.2934208}{{\em IEEE
  Transactions on Visualization and Computer Graphics}},
  \href{https://doi.org/10.1109/TVCG.2019.2934208}{26(1):321--331},
  \href{https://doi.org/10.1109/TVCG.2019.2934208}{Jan. 2020}.
  \href{https://doi.org/10.1109/TVCG.2019.2934208}
{doi: {{%
10\hspace{.1pt}\discretionary{.}{%
}{.}\hspace{.4pt}1109\discretionary{/}{%
}{/}TVCG\hspace{.1pt}\discretionary{.}{%
}{.}\hspace{.4pt}2019\hspace{.1pt}\discretionary{.}{%
}{.}\hspace{.4pt}2934208}}}


\bibitem{wingrave2006overcoming}
C.~A. Wingrave, Y.~Haciahmetoglu, and D.~A. Bowman.
\newblock Overcoming world in miniature limitations by a scaled and scrolling
  wim.
\newblock In {\em 3D User Interfaces (3DUI'06)}, pp. 11--16. IEEE, 2006.

\bibitem{woodburn_interactive_2019}
\href{https://doi.org/10.1109/VISUAL.2019.8933545}{L.~Woodburn, Y.~Yang, and
  K.~Marriott}.
\newblock \href{https://doi.org/10.1109/VISUAL.2019.8933545}{Interactive
  {Visualisation} of {Hierarchical} {Quantitative} {Data}: {An} {Evaluation}}.
\newblock \href{https://doi.org/10.1109/VISUAL.2019.8933545}{In {\em 2019
  {IEEE} {Visualization} {Conference} ({VIS})}},
  \href{https://doi.org/10.1109/VISUAL.2019.8933545}{pp. 96--100}.
  \href{https://doi.org/10.1109/VISUAL.2019.8933545}{IEEE},
  \href{https://doi.org/10.1109/VISUAL.2019.8933545}{Vancouver, BC, Canada},
  \href{https://doi.org/10.1109/VISUAL.2019.8933545}{Oct. 2019}.
  \href{https://doi.org/10.1109/VISUAL.2019.8933545}
{doi: {{%
10\hspace{.1pt}\discretionary{.}{%
}{.}\hspace{.4pt}1109\discretionary{/}{%
}{/}VISUAL\hspace{.1pt}\discretionary{.}{%
}{.}\hspace{.4pt}2019\hspace{.1pt}\discretionary{.}{%
}{.}\hspace{.4pt}8933545}}}


\bibitem{yalong_yang_maps_2018}
\href{https://doi.org/10.1111/cgf.13431}{Y.~Yang, B.~Jenny, T.~Dwyer,
  K.~Marriott, H.~Chen, and M.~Cordeil}.
\newblock \href{https://doi.org/10.1111/cgf.13431}{Maps and {Globes} in
  {Virtual} {Reality}}.
\newblock \href{https://doi.org/10.1111/cgf.13431}{{\em Computer Graphics
  Forum}}, \href{https://doi.org/10.1111/cgf.13431}{37(3):427--438},
  \href{https://doi.org/10.1111/cgf.13431}{June 2018}.
  \href{https://doi.org/10.1111/cgf.13431}
{doi: {{%
10\hspace{.1pt}\discretionary{.}{%
}{.}\hspace{.4pt}1111\discretionary{/}{%
}{/}cgf\hspace{.1pt}\discretionary{.}{%
}{.}\hspace{.4pt}13431}}}


\end{thebibliography}
\end{document}